\documentclass[12pt,preprint]{aastex}

\newcommand{\simgt} {\,\hbox{\lower0.6ex\hbox{$\sim$}\llap{\raise0.6ex\hbox{$>$}}}\,}
\newcommand{\simlt} {\,\hbox{\lower0.6ex\hbox{$\sim$}\llap{\raise0.6ex\hbox{$<$}}}\,}

\shorttitle{Environment and AGN feedback}
\shortauthors{Shin et al.}

\begin{document}

\slugcomment{Submitted to ApJ}

\title{Environmental effects on the growth of super massive black holes and AGN feedback}

\author{Min-Su Shin}
\affil{Princeton University Observatory, Peyton Hall, Princeton, NJ 08544-1001, USA}

\author{Jeremiah P. Ostriker}
\affil{Princeton University Observatory, Peyton Hall, Princeton, NJ 08544-1001, USA}
\affil{Institute of Astronomy, University of Cambridge, 
Madingley Road, Cambridge CB3 0HA, UK}

\author{Luca Ciotti}
\affil{Department of Astronomy, University of Bologna, via Ranzani 1, 
I-40127, Bologna, Italy}

\begin{abstract}
We investigate how environmental effects by gas stripping alter 
the growth of a super massive black hole (SMBH) and its host galaxy evolution, 
by means of 1D 
hydrodynamical simulations that include both mechanical and radiative 
AGN feedback effects. By changing the truncation radius of the gas distribution ($R_{t}$), 
beyond which gas stripping is assumed to be effective, 
we simulate possible environments for satellite and central galaxies 
in galaxy clusters and groups. 
The continuous escape of gas outside the truncation radius 
strongly suppresses star formation, while the growth of the SMBH is less affected by 
gas stripping because the SMBH accretion is primarily ruled by the density of the central region. 
As we allow for increasing environmental effects - the truncation radius decreasing 
from about 410 to 50 kpc - we find that the final SMBH mass declines from about ${\rm 10^{9}}$ to ${\rm 8 \times 10^{8} ~ M_{\odot}}$, 
but the outflowing mass is roughly constant at about ${\rm 2 \times 10^{10} ~ M_{\odot}}$. 
There are larger changes in the mass of stars formed, which declines from about ${\rm 2 \times 10^{10}}$ 
to ${\rm 2 \times 10^{9} ~ M_{\odot}}$, and the final thermal X-ray gas, which declines from about 
${\rm 10^{9}}$ to ${\rm 5 \times 10^{8} ~ M_{\odot}}$, with increasing environmental stripping. 
Most dramatic is the decline in the total time that the objects would be seen as quasars, which 
declines from 52 Myr (for $R_{t} = 377$ kpc) to 7.9 Myr (for $R_{t} = 51$ kpc). 
The typical case might be interpreted as a red and dead galaxy having episodic cooling flows followed by 
AGN feedback effects resulting in temporary transitions of the overall galaxy color from red to 
green or to blue, with (cluster) central galaxies spending a much larger fraction of their time in 
the elevated state than do satellite galaxies. 
Our results imply that various scaling relations for elliptical galaxies, 
in particular, the mass ratio between the SMBH and its host galaxy, 
can have dispersions due to environmental effects such as gas stripping. 
In addition, the simulations 
also suggest that the increase in AGN fraction in high-redshift galaxy clusters might 
be related to environmental effects which shut down the SMBH mass accretion in satellite galaxies and 
reduce their AGN activity.
\end{abstract}


\section{Introduction}

The role of the environment in galaxy evolution has been suggested in various forms 
which strip out gas from a galaxy 
and goes all the way back to the early suggestion by \citet{spitzer51}. For example, 
the ram pressure of the intracluster medium is a possible way to strip out 
gas from falling galaxies partially or completely in galaxy clusters 
and to stop 
the supply of cold gas for star formation \citep{gunn72,larson80,takeda84,gaetz87,begelman90,abadi99,
domainko06,tonnesen08,kapferer09}. Other 
possible processes include thermal evaporation and 
viscous stripping \citep{cowie77,livio80,nulsen82,nepveu85,valluri90,roediger08}. The combined 
gas loss by these different types of destruction effects is expected 
in galaxy clusters or groups \citep{stevens99,quilis00,toniazzo01,kawata08,
mccarthy08,smith09}. Tidal stripping can also 
play an important role in changing the gas and stellar mass of cluster or group galaxies 
\citep{merritt83,moore99,dercole00}.

Recent multi-wavelength observations have proved the loss of gases 
in cluster or group elliptical galaxies. In some elliptical galaxies, 
diffuse X-ray emitting gases 
show a long tail structure which can be explained by 
ram pressure stripping \citep[e.g][]{kim08a,randall08}. Infrared observations also revealed 
dust emissions that can be from gas stripped from elliptical galaxies 
\citep[e.g.][]{white91}.

A central galaxy which hosts satellite galaxies in groups and clusters
\footnote{A primary galaxy accreting satellites in group and cluster environment 
corresponds to the brightest cluster galaxy (BCG) in this paper. For isolated galaxies, 
the term {\it central galaxy} is correspondent to a primary galaxy which hosts 
satellites \citep[e.g.][]{ann08}. In this paper, the key feature to define satellites is that they 
experience gas stripping which is strong enough to affect their evolution.} 
exhibits different features compared to satellites which can be affected by the 
various stripping processes. Because the central galaxy sits near the bottom of a deep 
gravitational potential well, its hot gas halo is 
much larger than those of satellites \citep[e.g.][]{sun05}, 
and it is not surprising that it shows signatures of cooling flows in some cases 
\citep[see][for a review]{fabian94,reiprich04}.

Quite obviously, the differences in the stellar populations 
between satellites and central galaxies also have been 
studied, being relevant to our understanding of the environmental effects. For example, 
red satellite galaxies are redder than central galaxies of the same stellar mass 
\citep{vandenbosch08}. Many central galaxies in clusters display either recent star formation or 
ongoing star formation which may be related to the cooling flow from their hot gas halos 
\citep{cardiel98,rafferty06,bildfell08,odea08,pipino09} and which would be of reduced 
significance for satellite galaxies. Satellite galaxies in galaxy clusters seem to gradually 
lose of their gas by environmental effects and to truncate star formation, comparing 
their current and past star formation with those of field galaxies \citep[e.g.][]{balogh99}.

Differences in the properties of a central super massive black hole (SMBH) 
and its activity are expected between central and satellite galaxies, 
when considering the strong correlation between spheroidal galaxies and their SMBHs. 
The ratio of the SMBH mass over the spheroidal mass is found to be about 
$10^{-3}$ with a small dispersion for a large mass range in local galaxies \citep{kormendy95,magorrian98,
ferrarese00,gebhardt00,ferrarese02,yu02,marconi03,haring04}. If the 
growth of stellar mass and SMBH mass are coupled to the same kind of environmental effects, 
it is natural to conclude that the properties of SMBHs in satellite galaxies must be 
somewhat different 
from those of central galaxies. As we emphasized in our previous papers 
(Ciotti et al. 2009a, hereafter Paper I; Shin et al. 2010, hereafter Paper II) the self-regulated growth 
of SMBHs is profoundly dependent on how the accretion energy is converted to heat the ambient 
interstellar medium and how frequently and quickly the feedback process is ignited in the 
right place: we called these two key issues {\it the problem of energy conversion} 
and {\it the timing problem}. One would expect that the environmental 
effects might induce differences in the frequency of AGN feedback and the growth of the SMBHs by 
varying the conditions of the self-regulation process.

In this paper, we tackle the issue of environmental effects on the coevolution of 
the SMBH and its host galaxy by simulating the evolution of a galaxy in 
hydrodynamical models with a simplified setup of 
different environments. Although there is no current systematic investigation of the SMBH mass and 
AGN activity for separating 
satellite and central galaxies, the theoretical prediction from our simulations 
can be used to constrain the hypothesis of the coevolving SMBH and its host galaxy and 
the connection to environmental effects.

This paper is organized as follows. In Section \ref{sec:sim}, we describe the models and 
the simulations. The results are presented in Section \ref{sec:result}. Discussion 
and conclusions follow in Section \ref{sec:discussion}.

\section{Simulations}\label{sec:sim}

We adopt as a basic model the ${\rm B3^{w}}$ model that is given in \citet[][hereafter Paper III]{ciotti10}. 
This model incorporates 
both radiative and mechanical feedback effects from the central SMBH in 1D hydrodynamic simulations. 
The basic input physics is fully described in Paper I  and II. 
Here, we summarize the main ingredients of simulations and explain how to 
implement different environments.

We setup an initial galaxy following observed properties of local elliptical galaxies. 
First, the stellar mass is distributed to be consistent with 
Faber-Jackson relation and the Fundamental Plane, assuming 
an identical amount of stellar and dark matter mass 
within the half-mass radius \citep[e.g.,][]{tortora09}; all the relevant dynamical and structural 
properties of the models are given in Paper III. 
Second, the initial mass of the central SMBH 
$M_{\rm BH}$ is 0.001 of the initial stellar mass $M_{*}$. In all simulations, we set $M_{*} ~=~ 
2.9\times10^{11}\ M_{\odot}$, the effective radius $R_{e}$ is 6.9 kpc, and the central aperture 
velocity dispersion is 260 km/s (see Appendix for different initial velocity dispersions and their effects). 
Therefore, in our simulations, we are tracing the late evolution 
of ellipticals after the initial epoch of formation, i.e., of major growth of both the stellar 
component and of the central SMBH.

In addition to passive stellar feedback effects, i.e. mass losses and type Ia supernovae, 
the simulations include both 
radiative and mechanical feedback due to the SMBH as well as secondary stellar feedback effects 
from recurrent star formation. 
As discussed in Paper I, II, and III, the observational constraints such as 
the mass ratio between the SMBH and its host galaxy, the X-ray luminosity from 
hot diffuse gas, the quasar lifetime, and the recent star formation fraction, 
are better matched by simulations which include both feedback modes than 
simulations restricted to purely radiative or purely mechanical feedback models 
\citep[see][for a review of various forms of feedback]{begelman04}. 
Radiative feedback affects the ambient 
medium around the SMBH via 
radiative heating mediated by photoionization, Compton heating, and radiation 
pressure. These can be effective forms of AGN feedback over a large spatial range 
depending on a frequency range of AGN radiation 
\citep{sazonov04,sazonov05,fabian06,ciotti07}. Meanwhile, the typical form of mechanical 
feedback is represented by the 
interaction of nuclear winds and jets from accreting SMBHs which ultimately heat up 
the surrounding interstellar medium with maximum effectiveness in $r \ll 0.1$ kpc 
\citep{tabor93,binney95,begelman04,veilleux05,springel05a,konigl06,johansson09}.

In our models, we control the strength of radiative feedback with the parameter $\epsilon_{0}$ which is 
related to 
the radiative efficiency $\epsilon_{\rm EM}$ that determines 
the bolometric accretion luminosity from the SMBH 
\begin{equation}
L_{\rm BH} = \epsilon_{\rm EM} \dot{M}_{\rm BH} c^{2}
\end{equation}
by
\begin{equation}
\epsilon_{\rm EM} = \epsilon_{0} \frac{A \dot{m}}{1 + A \dot{m}}.
\label{eq:adaf}
\end{equation}
Equation \ref{eq:adaf} is a phenomenological implementation of the main features of the ADAF (Advection Dominated Accretion Flow) 
accretion \citep{narayan96} with the rescaled SMBH accretion rate with respect to the Eddington accretion rate 
$\dot{m} = \dot{M}_{\rm BH} / \dot{M}_{\rm Edd}$. The parameter $A$ is 100 as in 
Paper I and II. We use $\epsilon_{0} = 0.1$, which is favored by observational 
constraints for luminous quasar phase, i.e. when  $\dot{m}$ is high \citep[e.g.][]{soltan82} 
\footnote{In Paper III, we also present models with $\epsilon_{0} = 0.2$.}.

The parameter $\epsilon_{\rm w}^{\rm M}$ handles the mechanical efficiency of the AGN wind which is 
coupled to the bolometric accretion luminosity of AGN and affects the mass loss rate of gas by the wind 
\citep{kurosawa09a}. 
After the introduction of the normalized accretion luminosity with respect to the Eddington luminosity $L_{\rm Edd}$, 
\begin{equation}
l \equiv \frac{L_{\rm BH}}{L_{\rm Edd}} = \frac{A \dot{m}^{2}}{1 + A \dot{m}},
\end{equation}
the mechanical efficiency of the AGN wind $\epsilon_{\rm w}$ is
\begin{equation}
\epsilon_{\rm w} \equiv \frac{3 \epsilon_{\rm w}^{\rm M}}{4} \frac{l}{1 + 0.25 l}.
\end{equation}
In this paper, we fix $\epsilon_{\rm w}^{\rm M} = 3 \times 10^{-4}$ which is one of the explored models 
and suggested as the probable value 
in Paper III. The peak mechanical feedback efficiency $\epsilon_{\rm w}^{\rm M}$ 
governs how much kinetic energy, momentum, and 
mass has to be deposited to the broad-line region, with the mass loss coefficient of the AGN wind 
\citep{kurosawa09b,kurosawa09c}
\begin{equation}
\eta_{\rm w} \equiv \frac{3 \eta_{\rm w}^{\rm M}}{4} \frac{l}{1 + 0.25 l}.
\end{equation}
where $\eta_{\rm w}^{\rm M} = 1800 \epsilon_{\rm w}^{\rm M}$ which is consistent with observational 
constraints on the velocity of the broad-line regions and the ionized absorption outflows in quasars 
\citep[e.g.][]{krongold07}.

The important feature of our simulations is the boundary condition that mimics different environments having 
different strengths of gas stripping. In practice, this is accomplished by changing the 
position of the last grid point $R_{t}$ (hereafter called "truncation radius") 
where outflow boundary conditions are imposed. 
We assume that gas escapes beyond the truncation radius as various destruction processes caused by the environment 
(principally, ram-pressure stripping) 
reduce the size of hot gas halo, but do not totally destroy it 
\citep{brighenti99,sun05}. There is no infall of gas from outside these radii. 

The efficiency of gas stripping depends on both galaxy and intracluster medium properties 
\citep[see][for discussions]{hester06}. 
The gravity from mass distribution in the galaxy competes with ram pressure 
which depends on the density of intracluster medium and relative velocity of galaxies with 
respect to intracluster medium. Therefore, the stripping strength can vary significantly 
depending on the complex combinations of galaxy models and the flows of intracluster medium 
toward galaxies, i.e. the orbits of galaxies and the properties of the intracluster medium.

Even though this setup of different environments is too simple and the symmetric 
stripping never happens in real conditions, it will help us to 
understand how environmental effects might affect galaxy evolution with AGN feedback. Simply, small 
values of $R_{t}$ correspond to the satellite-like environment where gas stripping can be effective, 
while large values of $R_{t}$ 
represent the environment of central galaxies which can keep a large amount of diffuse hot gas, practically 
without any substantial loss of gas (but without the confining effect of some external pressure, as for example, 
the intracluster medium of the center of a galaxy cluster).

We test ten different truncation radii for the same initial stellar mass and SMBH mass. As given in Table 
\ref{tab:radius}, $R_{t}$ ranges from about 51 kpc to 413 kpc. This range corresponds to 
about 7 to 60 $R_{e}$. Initially, all galaxies have the same density and temperature distribution of gas 
within the truncation radius. Therefore, 
when density and temperature distribution of hot gas is same in all cases at a certain time, 
the total X-ray luminosity 
emitted from the hot gas is expected to be lowest for the smallest truncation radius, i.e. Run 1. The spatial resolution 
of all runs is exactly the same despite the differences of the truncation radii, having different numbers of 
radial grids. The inner most grid point is placed at 2.5 pc from the center in all runs.

The simulation follows the evolution of a galaxy, which begins at the cosmic age 2 Gyr and stops at 14 Gyr, 
by solving the hydrodynamic equations of the gas, being supplemented by prescriptions for star formation and stellar 
feedback. Because the simulations begin at 2 Gyr, i.e. 
a redshift of $z \sim$ 3.2 for the LCDM cosmology with 
$\Omega_{\rm m} = 0.3$, $\Omega_{\lambda} = 0.7$, and $H_{0} = 70$ km/s/Mpc, it is assumed that 
a bulk of stellar mass is already established at the beginning epoch \citep{renzini06,vandokkum10}.

As explained in our previous papers, the simulations are not meant to reproduce the cosmological context. 
Instead, we explore the physical link between the local small-scale physics around SMBHs and the global scale 
of a single galaxy during late evolutionary phases. In general, it is already difficult 
to achieve this dynamic range with the intricate 
prescriptions of AGN feedback, so that 
cosmological gas infall or galaxy merger/accretion are not 
implemented in our simulations 
\citep{dimatteo08,letawe08}. This limitation must be considered when interpreting our simulation results. 
However, the recent observations suggest that 
the final stage of fueling onto the SMBH is controlled primarily by the self-regulation 
process instead of large-scale episodic 
effects such as galaxy mergers \citep{grogin05,kollmeier06,li08,gabor09,silverman09,reichard09}. Moreover, 
in this paper we focus on effects by gas stripping only.

\section{Results}\label{sec:result}

\subsection{Evolution of models}\label{subsec:evolution}

In all computed models, the growth of the new 
stars and the SMBH is strongly regulated by how 
frequently active phases of SMBHs occur. 
At early times, all models are characterized by the accumulation of recycled gas and its 
cooling which finally results in the formation of new stars and almost simultaneous SMBH growth, 
with a strongly intermittent activity 
as shown in Figure \ref{fig:time1}. The coincident activity of star formation and SMBH accretion is 
consistent with blue colors of quasar-hosting elliptical galaxies at low and 
high redshifts and the recent star formation in AGN hosts 
\citep[e.g.][]{jahnke04,sanchez04}.

The repetition of star formation episodes and the active SMBH 
phases is basically caused by self-regulation of AGN feedback, which heats up the cooling flow for 
star formation and suppresses fueling to the SMBH responding the cooling rate. 
The time intervals between the peaks of star 
formation rate $\dot{M}_{*}$ and the central SMBH growth rate $\dot{M}_{\rm BH}$ become 
progressively longer and longer, even though the precise timing of the repetition varies in 
each run. This delay is the result of the decline of $\dot{M}_{*}$ with cosmic time, 
and of the competition between heating and cooling which finally 
resurrect star formation and fueling onto SMBHs. 
The early frequent activities of forming stars and feeding SMBHs 
result in the substantial increase in stellar mass and SMBH mass within 
the first two Gyrs as recycled gas from stars is used to form new stars in the central 
dense regions.

The dominance of SMBH growth over star formation changes over time, depending on how 
quickly feedback from the accreting mass affects the ambient interstellar medium (ISM) 
around the SMBH. For example, we investigate the evolution of $\dot{M}_{\rm BH}$ 
and $\dot{M}_{*}$ in detail around 3.23 and 13.31 Gyr in Run 6. As shown in Figure 
\ref{fig:time1_exam}, the peaks of $\dot{M}_{\rm BH}$ in the early accretion phase exceed 
the Eddington accretion rate for some short period of time. This high accretion onto the SMBH finally produces a 
strong feedback effect which suppresses the 
star formation rate. The late SMBH accretion rate, however, is much 
lower than the Eddington accretion rate even in the peak phase, while 
$\dot{M}_{*}$ rises to $10^{3} M_{\odot}$/yr for a moment. The ratio of $\dot{M}_{*}$ 
to $\dot{M}_{\rm BH}$ in this late accretion is consistent with the observed ranges of 
low-redshift quasars as shown in Figure \ref{fig:time1_exam} \citep[e.g.][]{shi09}. In the late accretion, 
the ratio of $\dot{M}_{*}$ over $\dot{M}_{\rm BH}$ ranges from about 10 to 300 for most time when 
$\dot{M}_{\rm BH} > 1 M_{\odot}$/yr. This range is consistent with the observed range even though 
the observed range might be limited by samples of quasars. 
Further discussion of this feature will be given in the next section.

\begin{figure}[!t]
\plotone{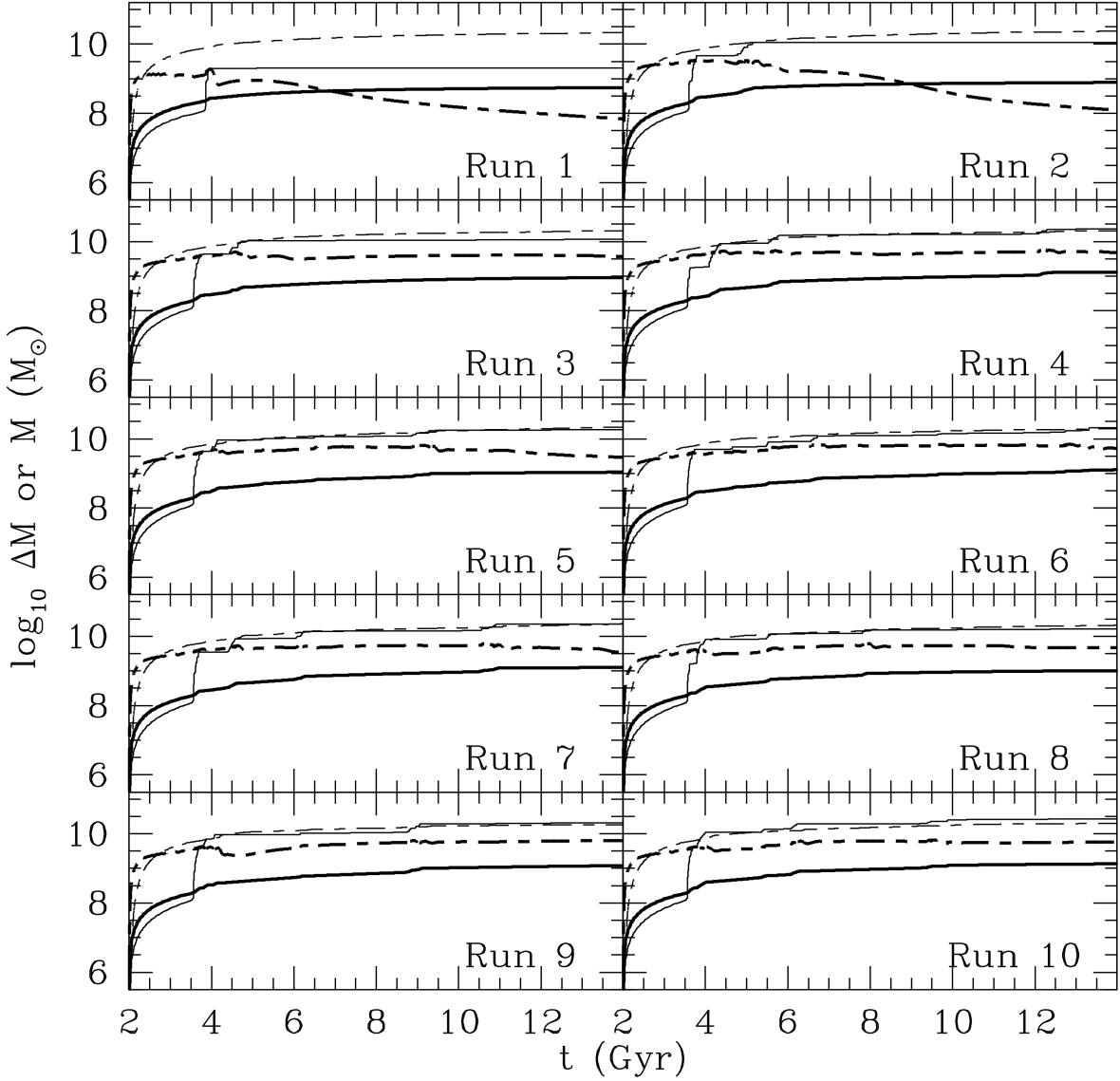}
\caption{Time evolution of mass components in the galaxy models. 
The different evolution of $\dot{M}_{\rm BH}$ and 
$\dot{M}_{*}$ for different values of $R_{t}$ results in different final values of the 
mass ratios between the SMBH and stars, and different amounts of gas that are blown out. 
The mass $\Delta {M}_{\rm BH}$ added to the SMBH ({\it thick solid line}) is in general much smaller than 
the change in the stellar mass, $\Delta {M}_{*}$ ({\it thin solid line}). 
The total amount of X-ray emitting hot gas contained in the galaxy ({\it thick short-long 
dashed line}) decreases as ${M}_{\rm BH}$, ${M}_{*}$, and 
the total mass of outflowing gas ({\it thin short-long dashed line}) increase, particularly 
in the models with small truncation radii.}
\label{fig:time2}
\end{figure}

As shown in  Figure \ref{fig:time2}, the mass of the hot, X-ray emitting gas in a galaxy 
does not change as much as the increase of the newly formed stellar mass or 
SMBH mass. We measure the total amount of the hot ISM within $10 R_{e}$ in all models 
except for Run 1 (where $R_{t}$ is smaller than $10 R_{e}$). The mass of outflowed gas is also 
measured at $10 R_{e}$. In the case of Run 1, these two quantities are instead measured at $R_{t}$. 
In general, stronger evolution is found in models with a small truncation 
radius. This has important consequences: for example, in Run 1 and 
2, the decrease in gas mass is so substantial that intensive star formation and 
SMBH accretion does not resume, even though heating by stellar and AGN 
feedback effects is not comparable to that in Run 9 and 10.

Importantly, the total amount of outflowing gas depends only weakly on the truncation radius 
(see Figure \ref{fig:time2}). 
Despite the small $R_{t}$, 
in Run 1 and 2 the total amount of the gas blown out is not higher than 
those of other runs because they have less frequent intensive star formation and 
SMBH accretion events which dump out energy and momentum to produce outflows. We examine this 
feature further in \S\ref{subsec:outflow}.

Figure \ref{fig:time_comp} represents the change of the mass budget as a function  
of $R_{t}$ at 4, 8, and 14 Gyr. The difference of the stellar mass appears 
at 4 Gyr, while other masses deviate less from each other. But as galaxies evolve, galaxies 
with smaller $R_{t}$ lack more X-ray emitting gas and stellar mass. It is naively true that the 
smaller galaxies have a smaller amount of gas and stellar mass because of their smaller $R_{t}$. 
But this trend is not very effective for the growth of SMBHs as we see in Figure \ref{fig:time_comp}, 
resulting in dispersions of the ratio between stellar mass and SMBH mass.  
$M_{BH} / M_{*}$ in Run 1 is about two times smaller than that in Run 10.

\subsection{AGN activity and feedback}\label{subsec:AGN}

\begin{figure}[!t]
\plotone{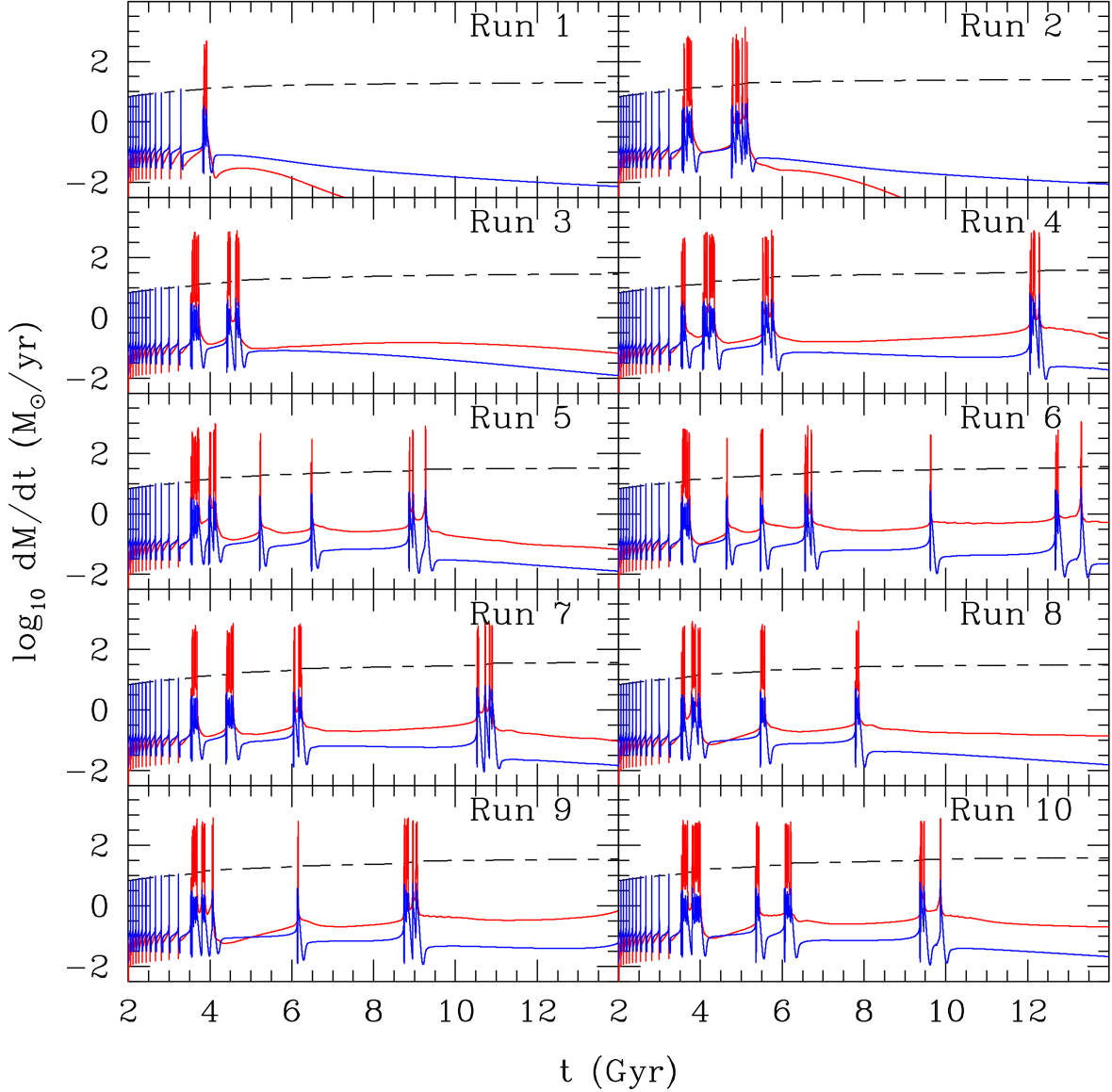}
\caption{SMBH mass accretion rate and star formation rate. 
The SMBH mass accretion rate $\dot{M}_{\rm BH}$ ({\it blue}) and star formation rate 
$\dot{M}_{*}$ ({\it red}) show a strong time-dependence with multiple intensive events 
consisting of several peaks. $\dot{M}_{\rm BH}$ is usually lower than the Eddington accretion rate 
({\it short-long dashed line}). Note how the frequency of the peaks of $\dot{M}_{\rm BH}$ and 
$\dot{M}_{*}$ depends on the truncation radius, particularly increasing from Run 1 ($R_{t} = 51.4$ kpc) 
to Run 3 ($R_{t} = 127.2$ kpc).}
\label{fig:time1}
\end{figure}

\begin{figure}[!t]
\plotone{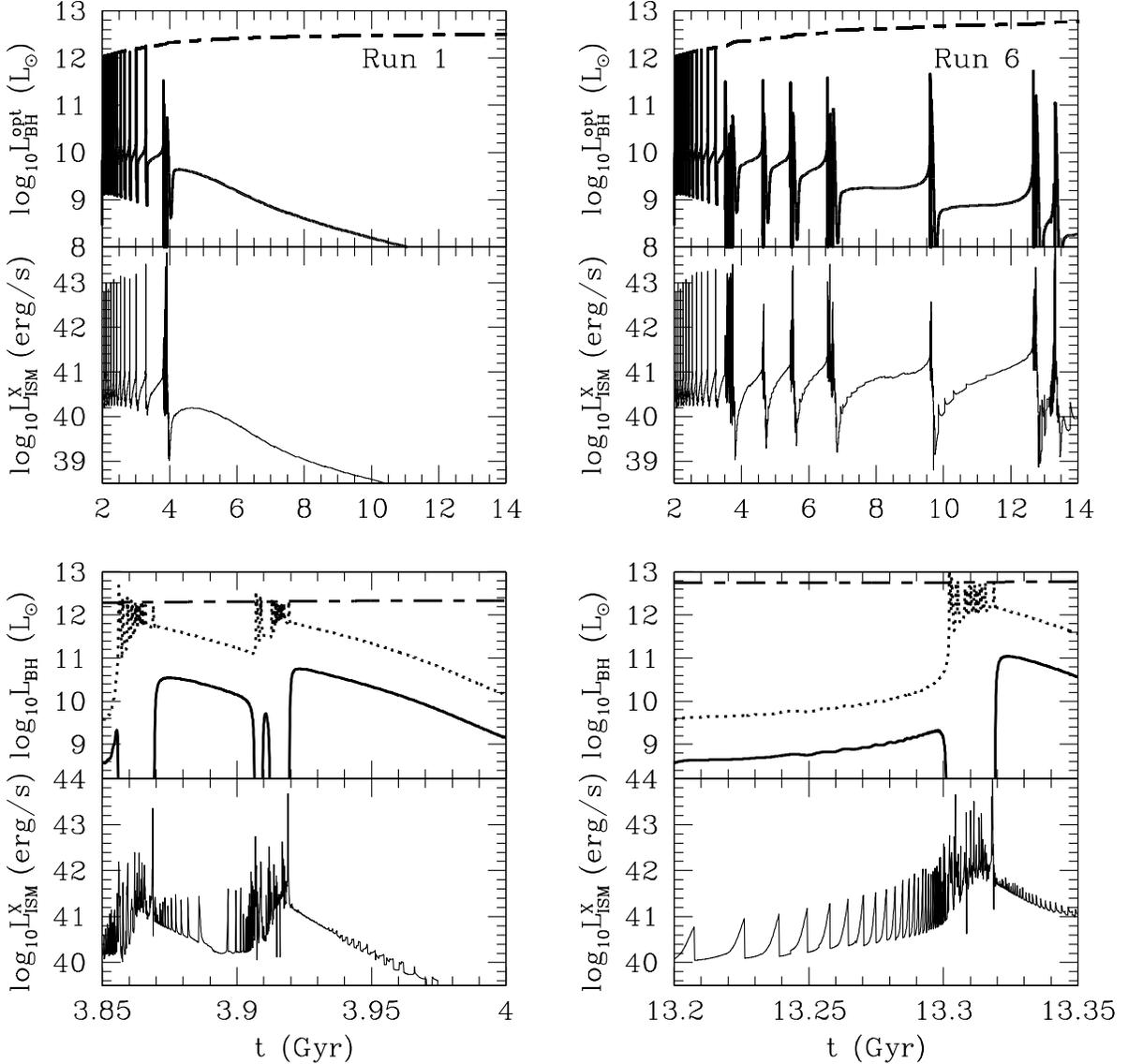}
\caption{
Optical luminosity from accretion onto SMBHs and X-ray luminosity from hot ISM 
in their host galaxies for Run 1 and 6. In both simulations, the optical luminosity of 
SMBHs $L_{\rm BH}^{\rm opt}$ ({\it thick solid line}) correlates with 
the change of X-ray luminosity $L_{\rm ISM}^{\rm X}$ 
({\it thin solid line}). The optical luminosity from the accretion is usually 
below the 10\% of the Eddington luminosity ({\it short-long dashed line}). 
We note that the optical luminosity is derived considering the extinction effects \citep{ciotti07}. 
The bottom plots show the complex evolution of Run 1 at about 3.9 Gyr and of Run 6 
at about 13.3 Gyr with bolometric luminosity from SMBH without dust extinction ({\it dotted line}). 
When $L_{\rm BH}^{\rm opt}$, i.e. the SMBH accretion rate, is high, 
$L_{\rm ISM}^{\rm X}$ shows the oscillatory structure because of feedback from 
the accreting SMBH. Yet, the accumulated cold gas finally strongly extincts optical luminosity from 
the accreting SMBH while $L_{\rm ISM}^{\rm X}$ and the SMBH bolometric luminosity 
increase around 3.86, 3.92, and 13.31 Gyr.
}
\label{fig:AGN_time}
\end{figure}

The frequency of AGN activity varies systematically with 
$R_{t}$ as 
shown in Figure \ref{fig:time1}. Smaller $R_{t}$ results less frequently in 
strong SMBH accretion events. Particularly, this change is significant for Runs 1 to 3. 
For example, Run 1 stops any further intensive 
accretion after 4 Gyr as shown in Figure \ref{fig:AGN_time}. Only early peaks of 
the SMBH optical luminosity are higher than the 10\% of the Eddington luminosity. 
However, Run 6 continues to shows AGN activity until 14 Gyr, while the late 
optical emission from the central SMBH is much lower than the 10\% of the 
Eddington luminosity as in Run 1 (see Figure \ref{fig:AGN_time}). 
In other runs such as Run 6 except Runs 1, 2, and 3, this late continuous AGN activity 
smooths out the dependence of AGN activity on $R_{t}$, and instead shows highly time-dependent 
stochastic effects. 
Remarkably, as we see in our simulations, 
recent observations show a general decrease in the ratio between the SMBH accretion 
rate and the Eddington rate as the redshift decreases \citep[e.g.][]{kollmeier06,ballo07}. 
In sum, AGN activity in satellite galaxies is expected to decrease with decreasing 
redshift much more than that of central galaxies.

We note that 
the late SMBH accretion can reach the limit of the Eddington accretion rate in models 
with only mechanical feedback effects (see Paper I and II). But in the current 
models with both radiative and mechanical feedback, the radiative pressure contribution 
due to dust opacity makes the late peaks of the accretion rate much lower than 
the Eddington rate \citep{ciotti07} in agreement with recent measurements of the 
accretion rate and hydrogen column density \citep{fabian09}. By 3.5 Gyr mass loss from 
the initial bulge accumulates an amount of cold gas which can be strongly affected by 
radiative feedback effects. Therefore, after about 3.5 Gyr all models do not have 
super-Eddington accretion phases as shown in Figure \ref{fig:time1}.

Strong AGN activity is generally correlated with the increase in 
X-ray luminosity from the hot ISM. Despite the differences in $R_{t}$ and the 
consequential difference in AGN activity, the response of the X-ray luminosity 
to the AGN activity is similar in Runs 2 and 6 as shown in Figure \ref{fig:AGN_time}. 
As the SMBH accretion rate increases, the oscillatory response of the hot ISM to AGN feedback 
begins to produce an oscillation of the X-ray luminosity. When the optical luminosity from the SMBH 
accretion temporally drops because of extinction, the X-ray luminosity quickly increases 
\citep[see][for a discussion on the evolution of the nuclear X-ray emission]{pellegrini09}.

\begin{figure}[!t]
\plotone{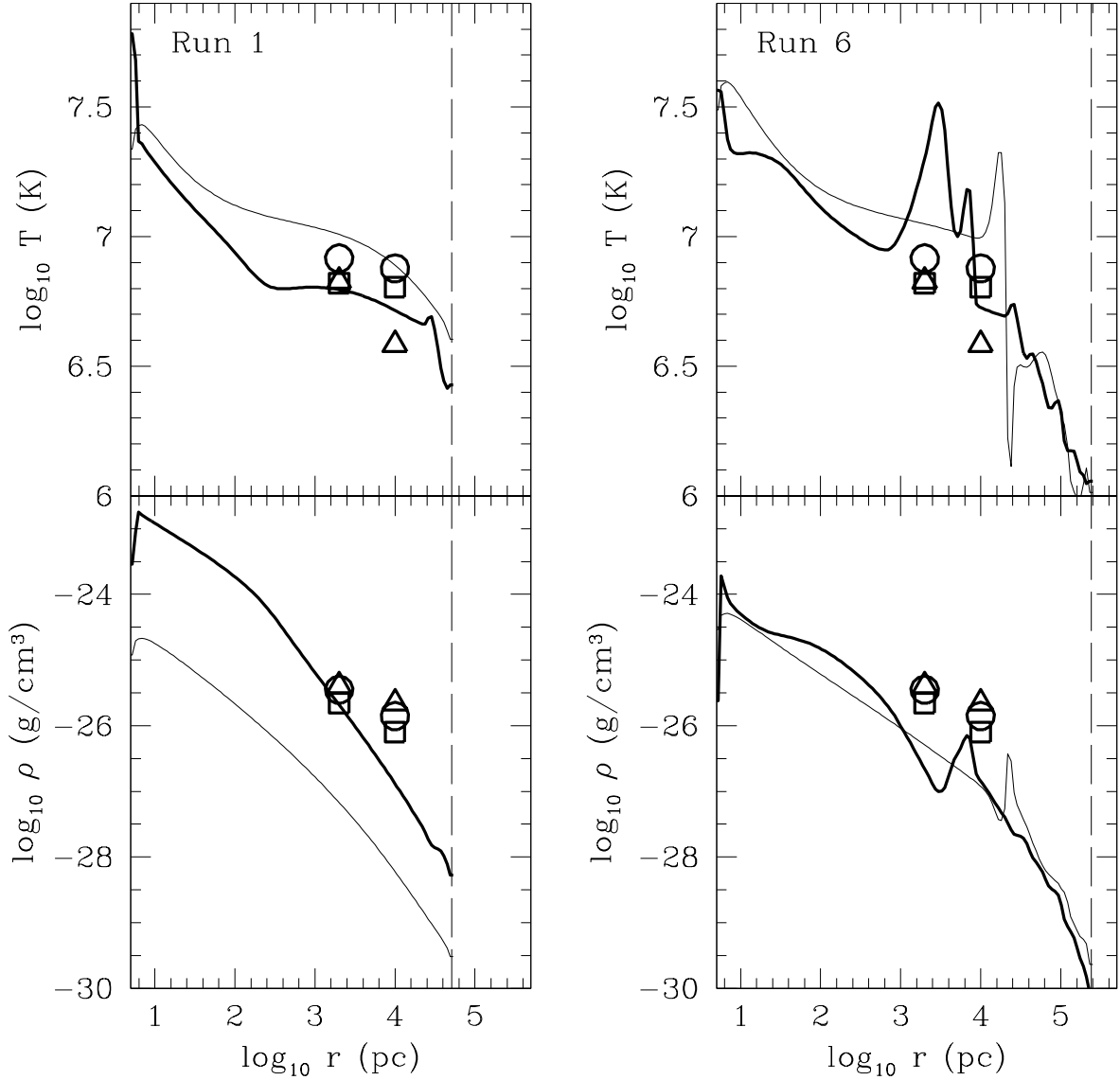}
\caption{
Radial distribution of the ISM temperature and density in Run 1 and 6. 
The truncation radii are indicated as dashed lines. 
At 14 Gyr 
({\it thin line}), Run 1 and 6 are found in the low-luminosity hot accretion 
phase; the higher central temperature and density in Run 6 are because of 
the influence of the larger final SMBH mass. 
However, Run 6 still shows the effect from its recent active phase 
forming the sharp shell structure. When both models are 
in the active phase at 3 Gyr ({\it thick line}) 
either increase of core temperature or shell structures appears in the 
radial distributions. Triangles, squares, and circles correspond to 
observed values at 2 and 10 kpc in NGC 4125, NGC 720, and NGC 6482, respectively, 
from \citet{humphrey06}.
}
\label{fig:AGN_space}
\end{figure}

The effects from AGN feedback can also be found in the radial distribution of gas temperature 
and density. Figure \ref{fig:AGN_space} shows the two cases of Run 1 and 6 at 3 and 14 Gyr. 
As we found in Figure \ref{fig:AGN_time}, both models are in an active phase at 3 Gyr. 
Basically, the active phase in both models causes the increase in the central temperature. 
In addition, the consecutive active phases leave the wiggle structure in both temperature 
and density distribution as the AGN feedback effects propagate to the 
outer regions on a sound-crossing time scale. 
At 14 Gyr, Run 1 is in the quite stationary, hot accretion phase, while Run 6 experienced 
a recent violent nuclear burst within 1 Gyr (see Figure \ref{fig:AGN_time}). 
Considering uncertainties of deprojection and radial binning effects in observation, the radial temperature and 
density distribution of Run 1 is within the observed range of local quiescent elliptical galaxies 
\citep[see][for examples of observation]{fukazawa06,humphrey06}, 
although its X-ray luminosity from the hot gas is lower than the observed range of luminosities 
for the same stellar mass. For example, the central density of the hot ISM is around $10^{-24}$ to $10^{-25} 
{\rm (g cm^{-3})}$ in all runs. But the recent AGN activity 
makes Run 6 have a more disturbed radial structure than Run 1. Moreover, we find a 
significant difference in the central density between the two runs at 14 Gyr.

\begin{figure}[!t]
\plotone{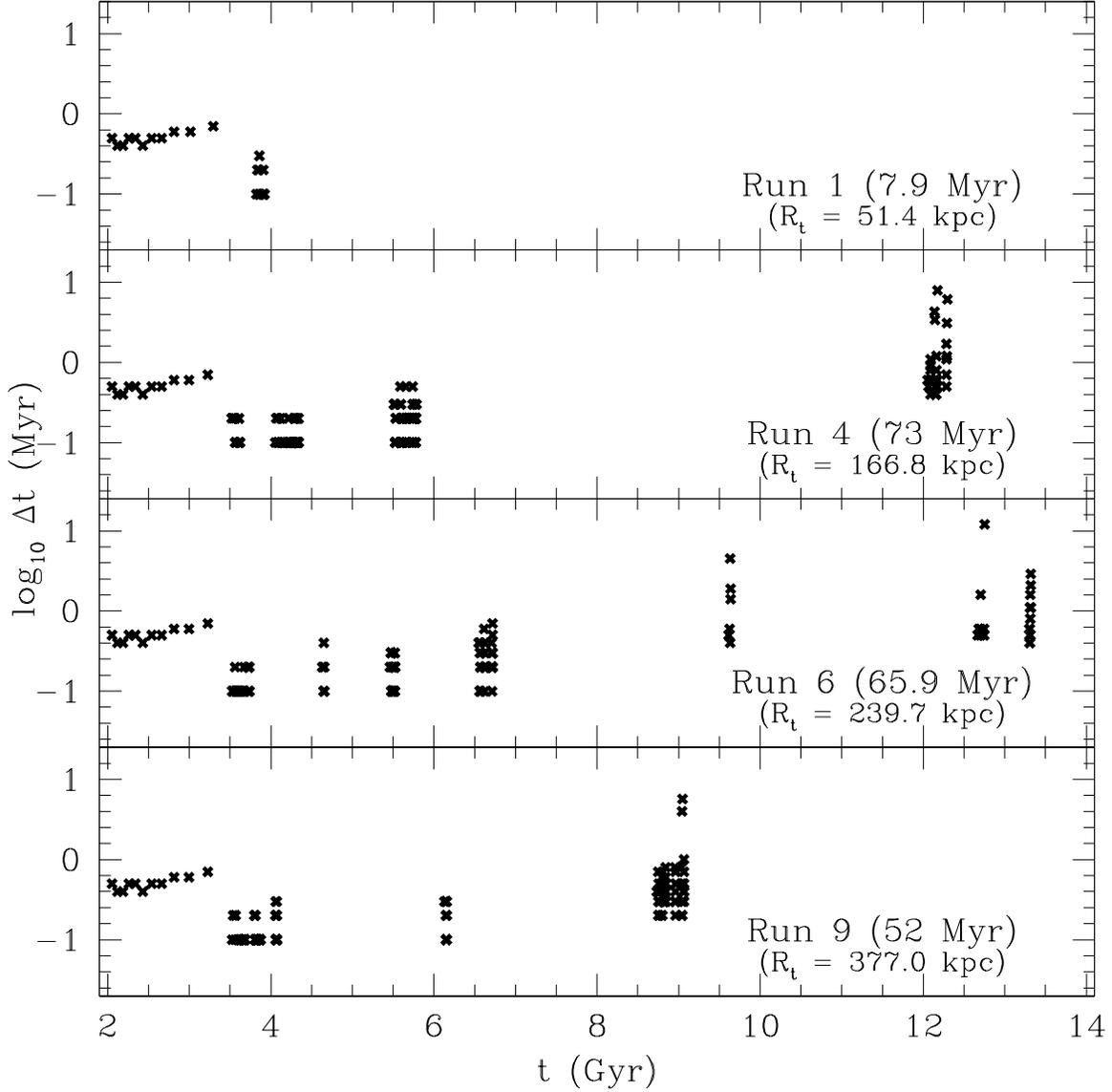}
\caption{
Episodic and net lifetime of optically luminous quasar phase (i.e. $M_{B} < -23$ mag) 
in Run 1, 4, 6, and 9. 
The episodic lifetime is presented for each peak of $\dot{M}_{\rm BH}$ 
as cross symbols. 
The net lifetime is given in the parenthesis 
next to the simulation number. The episodic lifetime increases as the simulation 
continues until about 3.5 Gyr when the peaks of SMBHs accretion rate are 
higher than the Eddington accretion rate.
}
\label{fig:AGN_lifetime}
\end{figure}

Finally, the episodic and net quasar lifetime is measured in all runs, defining the 
quasar phase with the optical luminosity from the SMBH in $B$-band $M_{B} < -23$ mag (in Vega 
magnitude) 
\citep[e.g.,][]{martini04} by using the typical spectral energy distribution of quasar in $B$-band 
\citep{elvis94}. This magnitude cut corresponds to about 
$L_{\rm BH} \sim 5\times 10^{45} {\rm erg ~ s^{-1}}$ with the median spectral energy distribution of 
quasars \citep{elvis94}. We define the quasar phase with this $L_{\rm BH}$ cut and effective bolometric 
luminosity from simulations. The effective bolometric luminosity represents bolometric luminosity from 
AGN with extinction by ISM as explained in Paper I. 
The episodic lifetime of quasar phase is an interval measured from the increase of the SMBH optical luminosity 
above $M_{B} < -23$ to the drop below this level, while the net quasar lifetime is the sum of 
these episodic lifetimes. Therefore, each peak of $\dot{M}_{\rm BH}$ has its episodic lifetime when it is 
above $M_{B} < -23$. 
In Figure \ref{fig:AGN_lifetime}, we present results for Runs 1, 4, 6, and 9. 
The episodic lifetime of each quasar phase varies significantly among 
the peaks of the SMBH accretion. In particular, an extremely short episodic quasar phase is 
found between long episodic bursts after about 3.5 Gyr. 
The episodic quasar lifetime ranges from about 0.1 Myr up to 
about 10 Myr. Even though the simulations with the large $R_{t}$ do not 
experience late luminous quasar phases within the last 2 Gyr (see Figure \ref{fig:time1}), 
the net lifetime of quasar phase generally increases as the $R_{t}$ increases. But, because 
the net lifetime in models do not include any possible active phase before 2 Gyr, i.e. 
the starting epoch of our models, a direct comparison of this net lifetime to 
observational constraints is not straightforward.

\subsection{Outflowing gas}\label{subsec:outflow}

Even though the boundary condition of our simulation might not be realistic 
in detail, it is useful 
to know the properties of outflowing gas. Probably, the gas blown out of the truncation 
radius of satellite galaxies will be mixed into the intergalactic medium although 
the gas will change its 
temperature and density after mixing with the intergalactic medium. As explained in 
\S\ref{subsec:evolution}, the outflowing mass is measured at $R_{t}$ for Run 1, while in other runs 
the mass flowed out at $10 R_{e}$, i.e. 69 kpc, is measured as the outflowing mass. 
As shown in Figure \ref{fig:time2}, the total amount of outflowing 
gas does not show a significant difference between the case of satellites and central 
galaxies, while the growth of stellar mass is quite low in Runs 1 and 2. 
Because models with a large $R_{t}$ do not lose much gas, 
the outflowing mass measured at $10 R_{e}$ represents the pollution of the central gas 
to outer regions where satellite galaxies can also contribute gas by stripping processes. 
Therefore, galaxies with a small $R_{t}$ 
are more efficient for enriching the intergalactic medium per galaxy.

\begin{figure}[!t]
\plotone{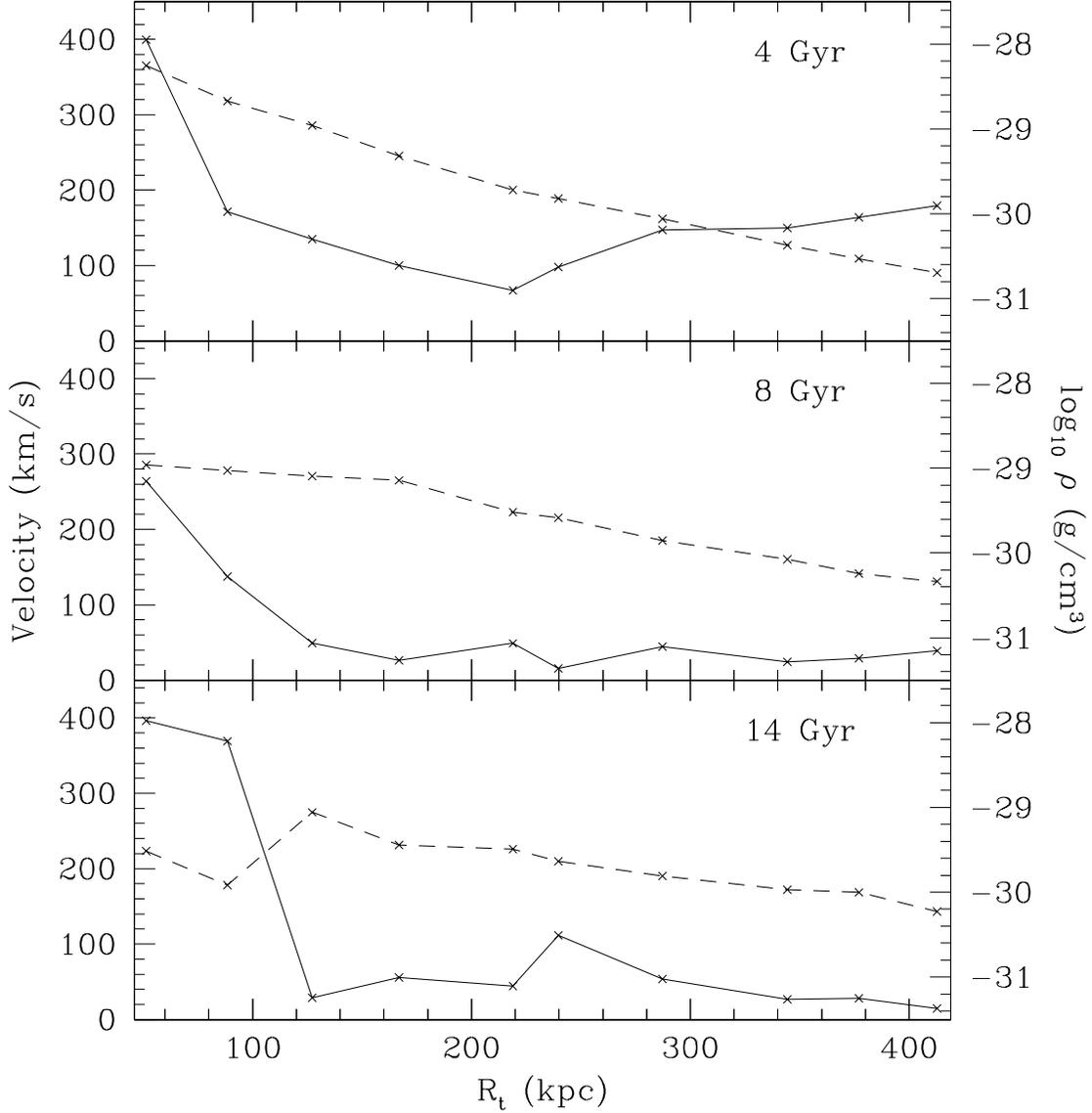}
\caption{
Outflow velocity and density as a function of $R_{t}$ at 4, 8, and 14 Gyr. 
Velocity ({\it solid line}) and density ({\it dashed line}) is estimated 
on the last radial grid point $R_{t}$ which represents the boundary of gas truncation. 
Models with small $R_{t}$ have the more dense and fast flows that pass out 
the truncation radius, producing the amount of mass loss which is comparable 
to that of models with the large $R_{t}$ when we measure the mass loss at $10 R_{e}$. 
We note that the mass flowing out at late times is negligible for large values of $R_{t}$.
}
\label{fig:outflow}
\end{figure}

Figure \ref{fig:outflow} illustrates how the high radial velocity and density of gas 
in models affects the total mass of outflowed gas. The galaxy with the smallest 
$R_{t}$ (Run 1) has expelled high-velocity dense gas continuously. Therefore, despite its 
smaller outflowing surface area at $R_{t} = 51.4$ kpc than other runs, its total 
mass of outflowed gas is substantial and comparable to the mass measured at $10 R_{e}$ in 
other runs. For example, the outflowing 
velocity for $R_{t} = 51.4$ kpc (Run 1) is close to 400 km/s at 4 Gyr. At 14 Gyr, 
simulations from Run 3 to 10 produce the low-velocity outflow which is slower than 100 km/s. 
But the difference in density becomes small at late time such as at 14 Gyr, as we also 
found in Figure \ref{fig:AGN_space}.

\subsection{Color of galaxies \label{subsec:color_gas}}

Color is one of the well observed properties of elliptical galaxies which 
are dominated by old stellar populations in the local universe. The simple old 
stellar population of elliptical galaxies implies that their star formation histories 
have been significantly quiet although a small amount of recent star formation is 
possible \citep[e.g.][]{lucero07,young09}. But environmental 
effects such as ram-pressure stripping are expected to make the color of satellite galaxies redder 
\citep[e.g.][]{martinez08,weinmann09}.

In order to investigate the spectral properties of our galaxy models, 
we construct a color-magnitude diagram of model galaxies by synthesizing spectra 
based on the star formation history of the simulations. 
Assuming solar metallicity or half solar metallicity for all stellar populations, the spectra are synthesized 
by using the BC03 model \citep{bc03}, and the corresponding magnitudes are estimated from the 
spectra with the SDSS filter bands \citep{fukugita96}. 
Because we do not consider the complicated metallicity distribution of stellar populations 
and dust extinction, the derived color cannot be directly compared with the observed 
colors of elliptical galaxies in various environments. Moreover, because 
colors of old stellar populations found in ellipticals are strongly affected by age effects with 
age - metallicity degeneracy \citep[see][for a review]{renzini06}, 
a direct comparison between our simulations with observed ellipticals need the 
careful consideration of age and metallicity distributions. Here, we just focus on 
how different star formation histories in different environments 
affect the color of galaxies in all runs. For the initial stellar mass at the beginning of simulations, 
we assumed that the same star formation history before 2 Gyr is described by 
\begin{equation}
\dot{M}_{*} \propto (t/\tau^{2}) ~ {\rm e}^{(-t/\tau)},
\end{equation}
where $t$ is a cosmic time and $\tau = 1$ Gyr \citep[e.g.,][]{nagamine06}. Therefore, the peak of the initial starburst 
is at $t = \tau$.

\begin{figure}[!t]
\plotone{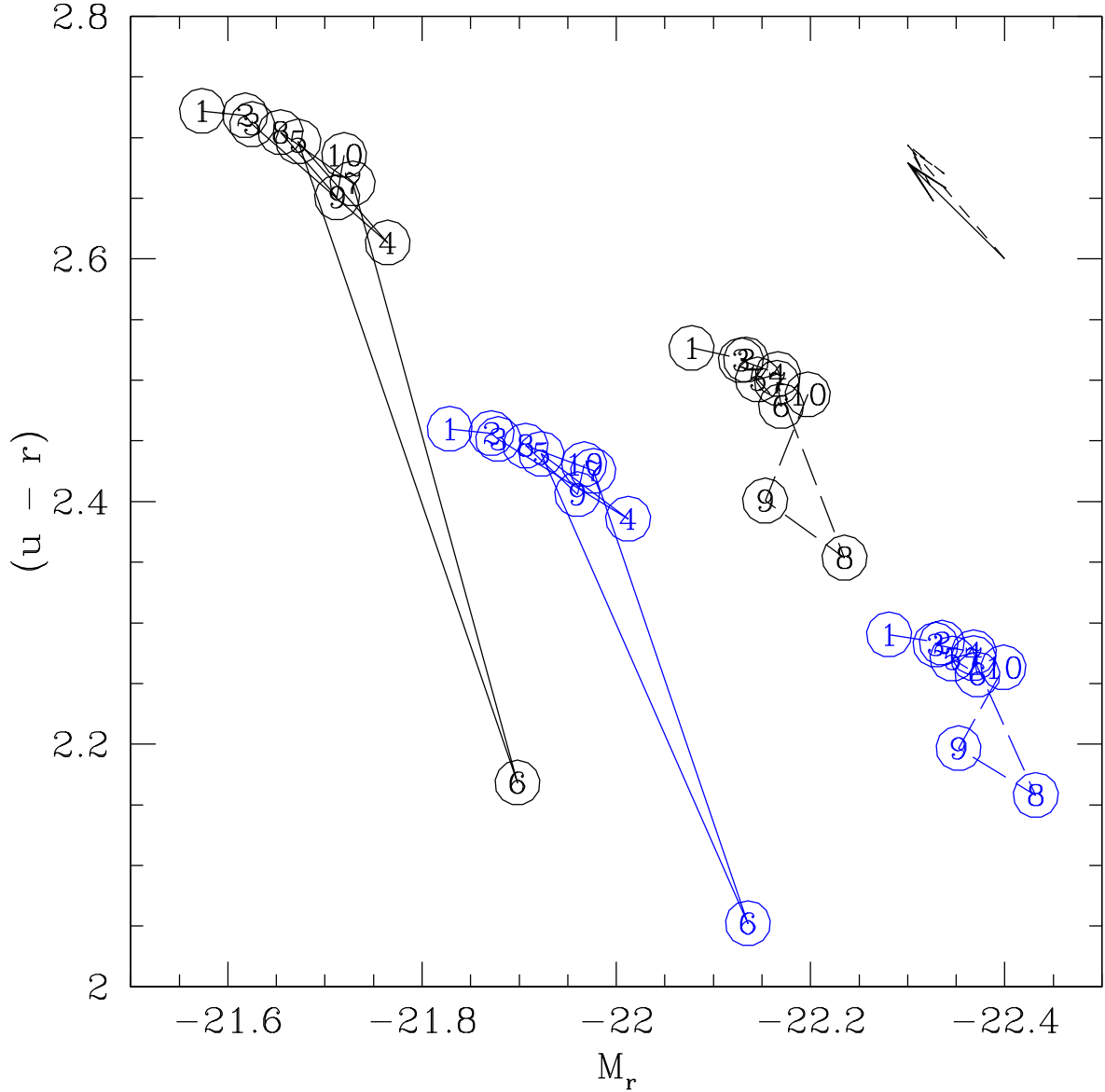}
\caption{Systematic color-magnitude diagram at 8.5 Gyr ({\it dashed line}) 
and 14 Gyr ({\it solid line}) without the optical radiation from the SMBH. 
From the star formation history 
of each run, we retrieve the synthesized spectra by using the BC03 model 
\citep{bc03}. Solar metallicity is assumed for black lines, while half solar 
metallicity is used for blue lines. The effects from dust extinction are not considered 
in the construction of spectra. The numbers in circles represent the name of the run. 
The arrows represents color excess for $A_{r} = 0.1$ with the dust extinction curves of the 
Milky Way ({\it solid line}) and Small Magellanic Cloud ({\it dashed line}) from 
\citet{pei92}, respectively.
}
\label{fig:CMD1}
\end{figure}

In Figure \ref{fig:CMD1}, we present the distribution of model galaxies in the SDSS 
$(u - r)$ rest-frame color and $r$-band absolute magnitude in AB magnitude 
at 8.5 and 14 Gyr. 
All galaxies are brighter and bluer at 8.5 Gyr than at 14 Gyr because 
most of the stellar mass is dominated by the initial old stellar population which changes its 
color by passive evolution. But there is a significant variation of the color for 
simulations with different truncation radii. For instance, 
Runs 8 and 9 at 8.5 Gyr show bluer color than other models, while becoming green 
galaxies temporally. At 14 Gyr, the color of the galaxy in 
Run 6 deviates strongly from those of other runs toward green color because 
of the recent star formation (see Figure \ref{fig:time1}). This 
temporary transition to green color with the increase in stellar mass 
can be a reason why moderate-luminosity AGN, i.e. a low-accretion AGN, 
are often found with a green color in the color-magnitude diagram \citep{hasinger08,schawinski09}.

\begin{figure}[!t]
\plotone{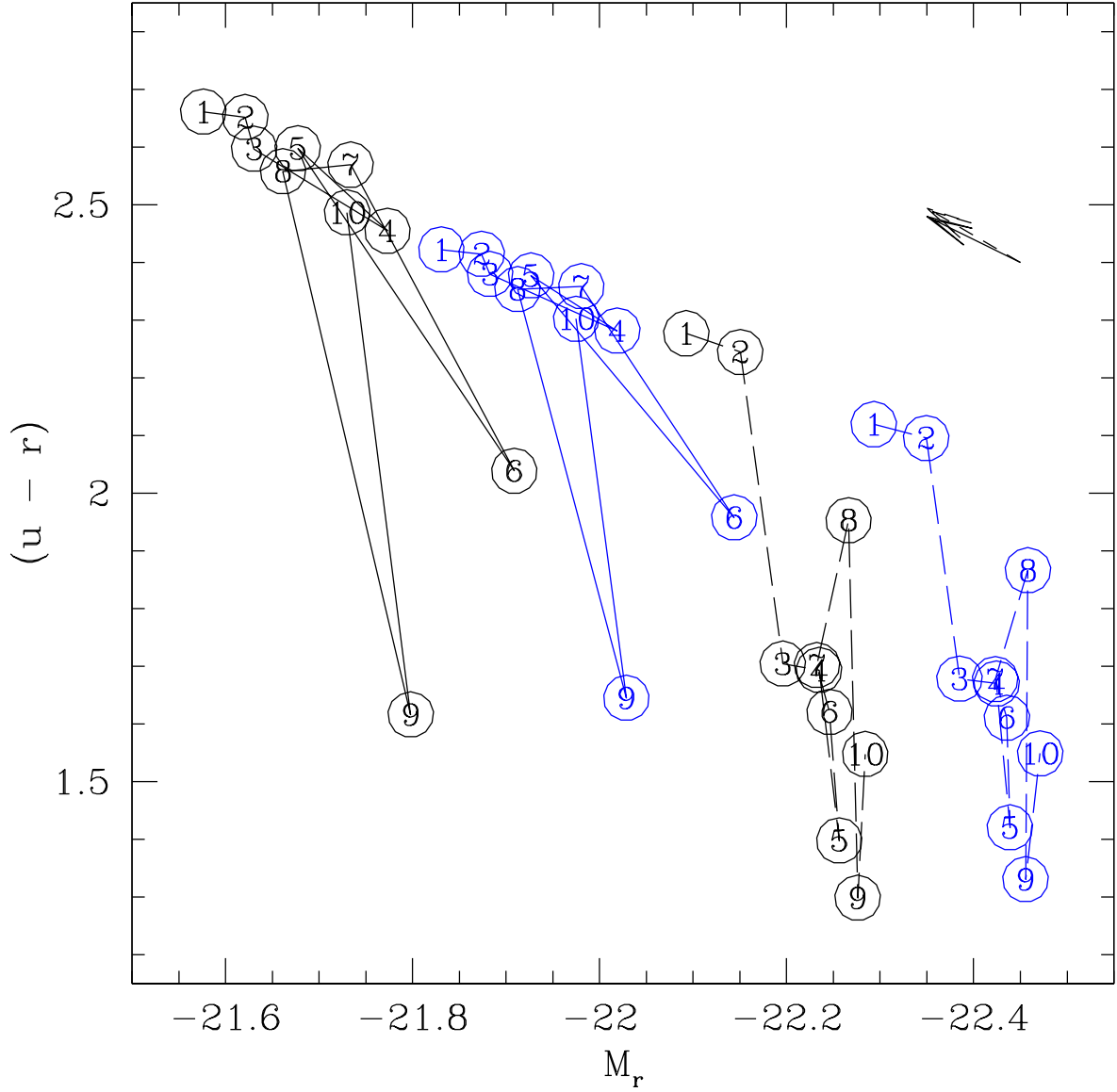}
\caption{Same as Fig. \ref{fig:CMD1}, but with the optical radiation from the SMBH. 
The included quasar spectrum is a luminosity-scaled spectrum of type-I quasar \citep{vandenberk01}. 
}
\label{fig:CMD2}
\end{figure}

In addition to the stellar flux, the color of galaxies can also be affected by AGN radiation if 
the latter is comparable or more luminous than the radiation from the host galaxy. 
To study this problem, we add the typical spectral 
energy distribution of type-I luminous quasars \citep{vandenberk01} to the synthesized stellar spectra 
of model galaxies. 
The quasar spectrum is scaled to the extinction-corrected optical luminosity 
from simulations, which correspond to the observed luminosity of quasars, before it is added. 
Because the quasar radiation is substantially bluer than the stellar radiation of quasar 
hosts, the addition of quasar radiation makes the global color of simulated galaxies bluer, 
as we show in Figure \ref{fig:CMD2}. Particularly, Run 9 is strongly affected by its quasar radiation. 
The almost coincident activity of star formation with SMBH accretion 
as shown in Figures \ref{fig:time1} and \ref{fig:time1_exam}, results in the shift of galaxy colors to the blue 
by radiation from young stars as well as that from quasar phase. Importantly, 
as observed in AGN host galaxies \citep[e.g][]{cidfernandes01,ammons09,silverman09}, 
the existence of blue young stellar populations is 
related to the strength of the SMBH accretion in our simulations. 
Therefore, Run 6 has the most significantly 
different color compared to other runs.


\section{Discussion and conclusions} \label{sec:discussion}

In this paper, we have investigated the effects of AGN feedback (in the form of combined radiative 
and mechanical energy and momentum deposition) on galaxies in different environments corresponding 
satellite and central galaxies.

Although not including fully realistic stripping effects, our simulations predict various observable 
features that can be compared to the current and 
future observations, and which should change in a systematic way between central and satellite galaxies. 
The simulation results need to be interpreted as a guide to understand a general consequence 
of environmental effects. Here, we discuss implications from our simulations.

First of all, the local scaling relationship such as the mass ratio between the SMBH and its host galaxy 
can have a dispersion due to environmental effects. For example, if gas in a galaxy has experienced stripping, 
the growth of its SMBH is less disturbed than star formation which generally happens over the whole galaxy. 
Hence, the mass ratio between the SMBH and the host galaxy of fixed galaxy mass 
could be significantly higher in satellite ellipticals in galaxy clusters. 
This also implies that a bias effect on deriving the scaling relationship 
probably exists due to selecting satellite galaxies less likely as observation samples. Objects measured for their SMBH mass are 
generally bright objects which might not include galaxies having experienced gas stripping. 
Even though the recent measurements 
of the SMBH mass might have an intrinsic 
scatter of the ratio \citep[e.g.][]{kim08b,gultekin09}, 
the current measurements of the SMBH mass in satellite elliptical galaxies have 
not been conducted systematically. Probably, the systematic investigation of early-type galaxies in 
the Virgo Cluster might be useful to test our results for satellites in future 
\citep{cote06,decarli07,gallo08}. But we warn that the uncertainty and biases in measuring the SMBH mass 
might dominate the dispersions in the mass ratios even for galaxies in the Virgo 
cluster \citep[e.g][]{bernardi07}.

The results of our simulations also imply the less frequent and 
earlier termination of AGN activity 
in satellite galaxies compared to central galaxies (i.e. galaxies 
with a large $R_{t}$). This result can 
be tested against the AGN fraction in galaxy clusters and groups \citep[e.g.][]{gilmour07}. 
The observed AGN fraction in galaxy clusters 
is generally higher than in field galaxies partially because of the high number density of galaxies (i.e. the 
effect from the cluster richness) and how AGN activities are selected in radio, X-ray, and IR observations 
\citep[e.g.][]{hickox09}. When considering this effect of the high number density of galaxies, 
AGN activity actually seems suppressed in galaxy clusters \citep[e.g.][]{koulouridis10}. 
The evolution of AGN fractions in galaxy clusters as a function of redshift also  
shows the increase in AGN activity at higher redshifts when satellite galaxies are still frequently active 
in their SMBH accretion, supporting our hypothesis of environmental effects on AGN activity \citep{galametz09,
martini09}. 
The frequent multiple intensive SMBH accretion events in simulated central galaxies is consistent with 
several distinctive observable features of cluster central galaxies. For example, many central galaxies show 
multiple outburst structures such as several gas clumps and surrounding ripples 
even though the origin of those observed structures and what kind of AGN activity causes those 
is still uncertain \citep[e.g.][]{fabian05,graham08,blanton09,
clarke09}.

The very weak dependence of the ejected mass on $R_{t}$ implies that 
satellite galaxies can be important sources of 
metal enrichment in the intracluster medium. Even though our simulation 
does not follow the evolution of the metal contents in the 
outflowed gas, the mass stripped out of the galaxy, which is about $2 \times 10^{10} {\rm M_{\odot}}$ (see Figure \ref{fig:time_comp}), 
must be metal-rich being produced by gas recycling from the evolving stars and supernovae explosion. 
As explained in \S\ref{sec:sim}, our models are similar to the {\it leaky box model} of galactic chemical evolution 
for satellites \citep{pagel09} because of the outflowing 
boundary condition. Finally, the outflows from both satellites and 
central galaxies might cause mixing which is coupled 
to AGN activity and turbulence within the intracluster medium \citep[e.g.][]{osullivan05}. The detailed 
abundance patterns in the cluster gas should thus reflect the outflow from the satellite 
cluster ellipticals in addition to the processes by a cluster central galaxy 
\citep{domainko06,schindler08,rasmussen09,sivanandam09}.

The recent star formation in cluster central galaxies \citep[e.g.][]{rafferty08,kirkpatrick09} 
is also naturally explained in our simulations by continuous 
recurrence of cooling flows and consequently heating by AGN feedback in galaxies with a large $R_{t}$. 
In our simulations, star formation from a cooling flow is temporally permitted while some amount of 
the cooling flow is accreted onto the SMBH and triggers AGN feedback which finally heats up and reverses 
the cooling flow as the cool gas is consumed by star formation. 
Contrary to a classical solution to the cooling flow problem in the cluster 
galaxies which stops the cooling flow completely \citep[see][for a review]{fabian94}, our solution to 
the cooling flow problem for the cluster central galaxies is simply permitting temporary rejuvenation 
of star formation and AGN activity which is self-regulated, leading to the termination 
of the cooling flow 
\citep[see][for a further discussion]{ciotti07}. This kind of solution can be called the 
{\it intermittent cooling flow} scenario \citep{salome06,bildfell08,pipino09,wilman09}. A further 
test of our results will be the Eddington 
ratio of the rejuvenated SMBH accretion in local cluster central galaxies because 
the SMBH accretion rate 
in the simulated central galaxies is significantly lower than the Eddington accretion 
rate despite the change of $R_{t}$.

To make our simulations more realistic, we need to improve our simulations by including two important physics: 
feedback by a radio jet from AGN activity and cosmological setup of environmental effects. In addition to 
the limits of our simulations already mentioned in our previous Paper II, for example, 1D calculations, and 
possible different initial conditions such as models presented in Appendix, 
the radio 
jet is an important feedback component for cluster central galaxies 
in environments of galaxy clusters and groups \citep{begelman04,konigl06}. 
The observation of radio jets proves that the kinetic energy of jets is enough to be an effective feedback mode 
even though we do not have a clear explanation about how much of the relevant energy is transported to ISM and 
intracluster medium \citep{mcnamara07,shin11}. 
We note that powerful narrow jets such as that seen in the giant elliptical 
M87 tend to drill through the ambient gas in the galaxy depositing some amount of energy with a 
relativistic fluid directly to the intergalactic medium \citep{ferrari98,owen00}. 
Our implementation of stripping effect from different environments is 
simply parametrized by $R_{t}$ in one dimension, 
and is not close to dynamically changing environmental effects in cosmological 
evolution \citep[e.g.][]{takeda84,vollmer01,toniazzo01}. Moreover, our simulations do not include effects from tidal gravitational field 
which is commonly expected in cluster or group environments \citep{dercole00}. 
Although the simple implementation makes the interpretation 
of simulation results 
obvious, the direct test of simulations will require better but complicated models of ram-pressure stripping and 
other environmental processes such as tidal stripping 
in galaxies which have realistic orbits in galaxy clusters and groups. 

\appendix

\section{Models with lower and higher velocity dispersions}

We also examine models with lower and higher initial stellar velocity dispersions of model galaxies than the  
fiducial model given in the main text. These models correspond to lower 
and higher ratio of stellar mass over SMBH mass compared with the observed ratio in the local universe \citep{ciotti09a,shin10,ciotti10}. 
Here, we summarize results of the models Run 2, 4, 6, 8, and 10 with the lower velocity dispersion 240 km/s and higher 
velocity dispersion 280 km/s than the fiducial velocity dispersion 260 km/s.

\begin{figure}[!t]
\plotone{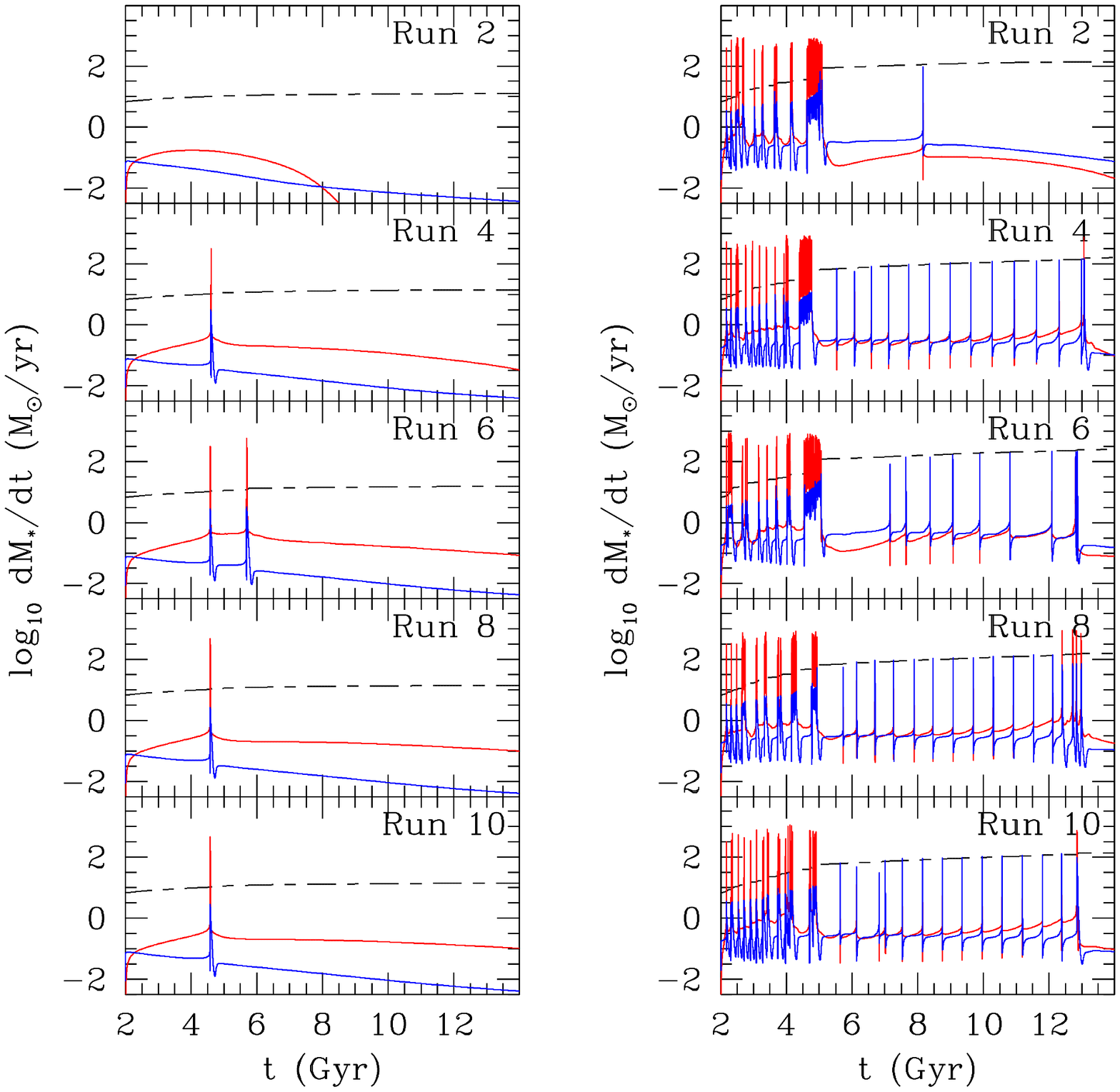}
\caption{Same as Fig. \ref{fig:time1}, but with the initial velocity dispersions of 240 km/s ({\it left}) and 
280 km/s ({\it right}) for Runs 2, 4, 6, 8, and 10.}
\label{fig:app1}
\end{figure}

As shown in Figure \ref{fig:app1}, the general trend of these models is not different from that of the fiducial model. 
One of the main differences is the frequency of the active SMBH accretion and star formation phase. As expected, models 
with the lower initial velocity dispersion experience the less number of active phases than the others because of a 
less amount of stellar mass, which supplies materials for SMBH accretion and star formation. The second important 
feature is a high intensity of star formation rate in the beginning of the models with the higher velocity dispersion, while 
showing a lower SMBH accretion rate at the same time than the fiducial models. Due to a larger stellar mass in these models, 
the initial accumulation of cooling gas is large enough to trigger the intensive star formation in the beginning. But this 
intensive star formation competes with SMBH accretion, resulting in the lower SMBH accretion rate. In the late stage 
of evolution, the models with the higher velocity dispersion also show higher SMBH and star formation rates than the fiducial 
models. In particular, the SMBH accretion rate is close to the Eddington accretion rate.

\begin{figure}[!t]
\plotone{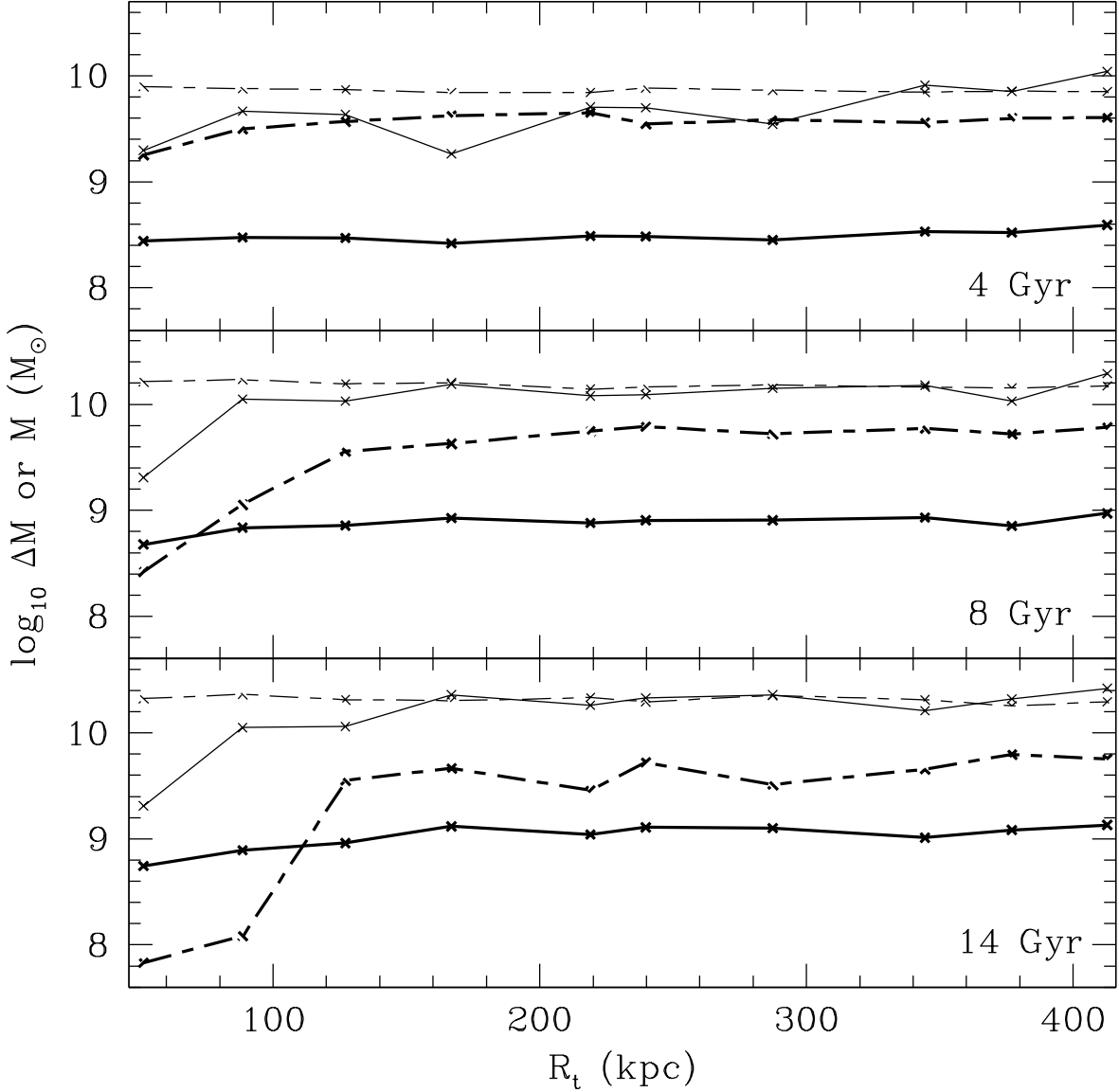}
\caption{Comparison of the mass budget at 4, 8, and 14 Gyr for different values of $R_{t}$. 
The presentation follows the same 
line styles as given in Figure \ref{fig:time2}. At 2 Gyr differences among the simulations are 
not significant. As effects from different star formation and 
SMBH accretion history accumulate, 
the increase in stellar mass and the total mass of gas in Run 1 and 2 begin to deviate from other 
simulations. But the amount of outflowing gas does not change much with respect to $R_{t}$.}
\label{fig:time_comp}
\end{figure}

\begin{figure}[!t]
\plotone{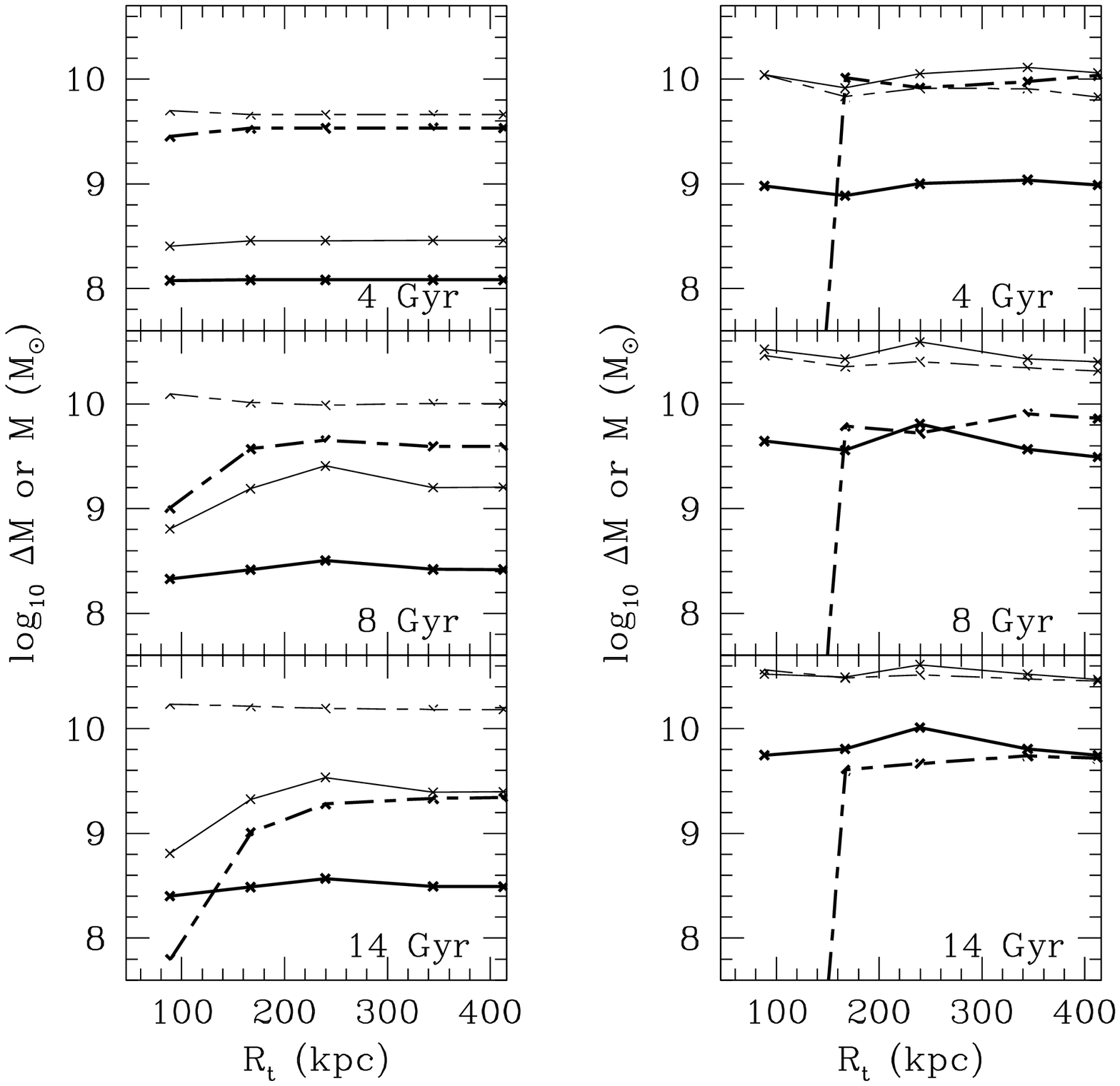}
\caption{Same as Fig. \ref{fig:time_comp},  but with the initial velocity dispersions of 240 km/s ({\it left}) and 
280 km/s ({\it right}) for Runs 2, 4, 6, 8, and 10.}
\label{fig:app2}
\end{figure}

Figure \ref{fig:app2} shows that the general patterns found in Figure \ref{fig:time_comp} are still valid with the 
lower or higher initial stellar velocity dispersion. But the change of ratio between stellar and SMBH masses in 
the models with the higher velocity dispersion is less dependent of the different truncation radii. This weak  
sensitivity to the environmental effect is caused by the strong and more late SMBH accretion and the 
dominating high star formation rate in the models with the higher velocity dispersion.

\acknowledgments

\begin{figure}[!t]
\plottwo{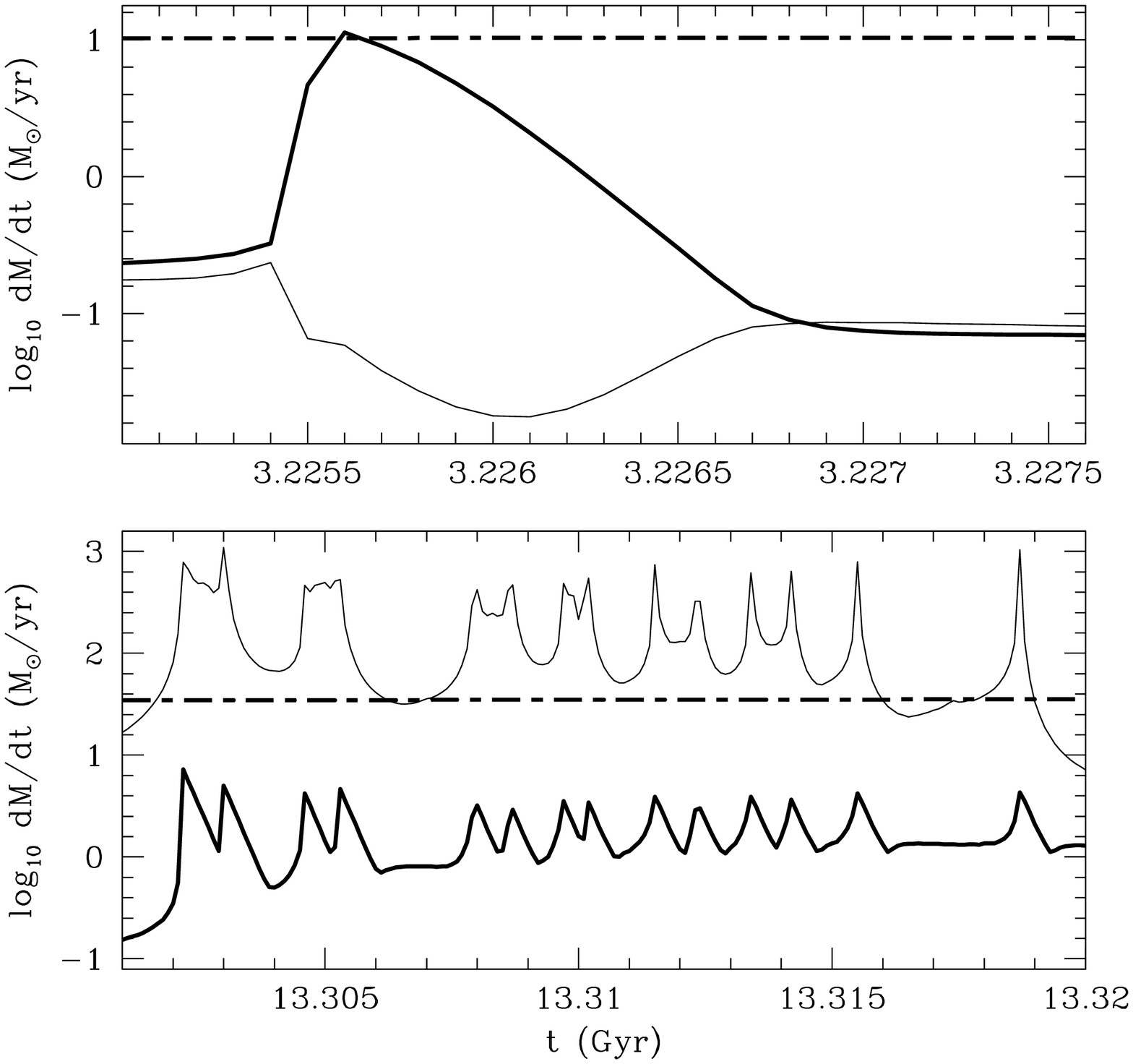}{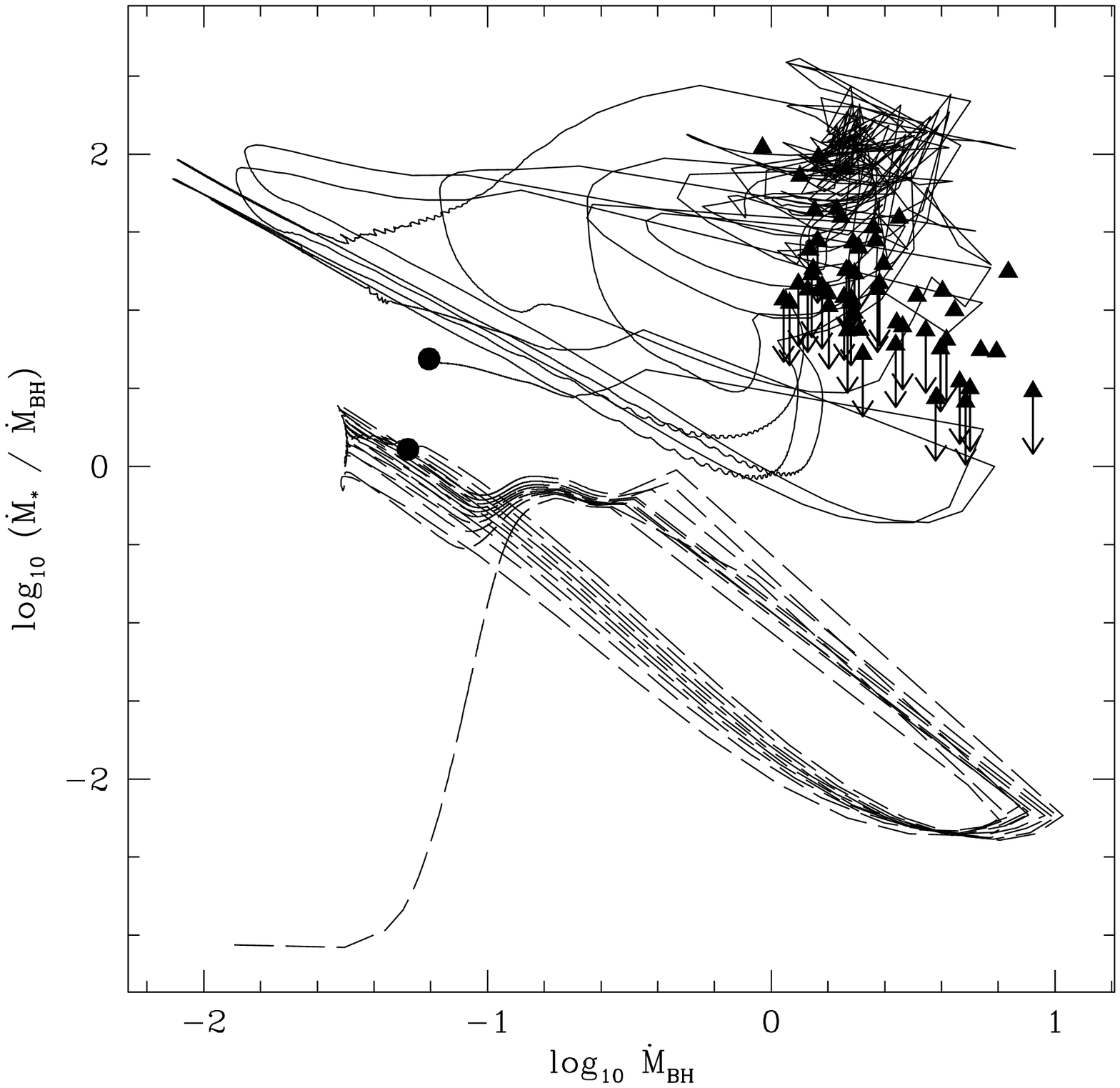}
\caption{Star formation rate ($\dot{M}_{*}$) and SMBH mass accretion rate ($\dot{M}_{\rm BH}$) 
around 3.23 Gyr ({\it left top}) and 13.31 Gyr ({\it left bottom}) and the distribution of 
$\dot{M}_{*}/\dot{M}_{\rm BH}$ with respect to $\dot{M}_{\rm BH}$ ({\it right}) in Run 6. 
Only early peaks of $\dot{M}_{\rm BH}$ 
({\it thick solid line}) are higher than the Eddington accretion rate ({\it thick short-long 
dashed line}) as shown in the left top panel. 
The early SMBH accretion is so strong that 
the following star formation is suppressed, while the late SMBH accretion 
is predominated by the concurrent intensive star formation.
When $\dot{M}_{*}$ ({\it 
thin solid line}) is high, $\dot{M}_{\rm BH}$ is generally low in the early SMBH accretion as apparent in the left top panel. 
But when $\dot{M}_{*}$ is high, $\dot{M}_{\rm BH}$ is also high in the late SMBH accretion. 
The difference between the early SMBH accretion pattern and the late pattern is 
clearly found in the right panel showing the change for the time earlier than 3 Gyr ({\it dashed line}) 
and later than 7.5 Gyr ({\it solid line}). Dots represent 3 and 7.5 Gyr, respectively. 
Triangles and triangles with arrows represent observed ratios and upper limits from \citet{shi09}, respectively.}
\label{fig:time1_exam}
\end{figure}

We are grateful to Michael Strauss, James Gunn, Gillian Knapp, Renyue Cen, Jenny Greene, 
and Christy Tremonti for useful discussions and careful 
reading. We thank the anonymous 
referee for comments which improved this manuscript. We also 
thank Yong Shi for the data used in Figure \ref{fig:time1_exam}. 
M.-S.S. is supported by the Charlotte Elizabeth Procter Fellowship of Princeton
University. Computations were performed on the computational facilities of 
PICSciE (Princeton Institute for Computational Science and Engineering).


\clearpage


\begin{table}[!t]
\caption{Truncation radius $R_{t}$ of the hydrodynamical grid}
\begin{tabular}{cc|cc} \hline \hline
Run & Radius (kpc) & Run & Radius (kpc) \\ \hline
1 & 51.4 & 6 & 239.7 \\
2 & 88.5 & 7 & 287.3 \\
3 & 127.2 & 8 & 344.4  \\
4 & 166.9 & 9 & 377.0 \\
5 & 218.9 & 10 & 412.8
\tablecomments{The effective radius of the galaxy models is 6.9 kpc at the beginning of the simulation.}
\end{tabular}
\label{tab:radius}
\end{table}

\begin{thebibliography}{}
\bibitem[Abadi et al.(1999)]{abadi99} Abadi, M.~G., Moore, B., \& Bower, R.~G.\ 1999, \mnras, 308, 947 
\bibitem[Ammons et al.(2009)]{ammons09} Ammons, S.~M., Melbourne, J., Max, C.~E., Koo, D.~C., \& Rosario, D.~J.~V.\ 2009, \aj, 137, 470 
\bibitem[Ann et al.(2008)]{ann08} Ann, H.~B., Park, C., \& Choi, Y.-Y.\ 2008, \mnras, 389, 86 
\bibitem[Balogh et al.(1999)]{balogh99} Balogh, M.~L., Morris, S.~L., Yee, H.~K.~C., Carlberg, R.~G., \& Ellingson, E.\ 1999, \apj, 527, 54 
\bibitem[Ballo et al.(2007)]{ballo07} Ballo, L., et al.\ 2007, \apj, 667, 97 
\bibitem[Begelman \& Fabian(1990)]{begelman90} Begelman, M.~C., \& Fabian, A.~C.\ 1990, \mnras, 244, 26P 
\bibitem[Begelman(2004)]{begelman04} Begelman, M.~C.\ 2004, Coevolution of Black Holes and Galaxies, Cambridge University Press, edited by L. C. Ho, p. 374
\bibitem[Bernardi et al.(2007)]{bernardi07} Bernardi, M., Sheth, R.~K., Tundo, E., \& Hyde, J.~B.\ 2007, \apj, 660, 267 
\bibitem[Binney \& Tabor(1995)]{binney95} Binney, J., \& Tabor, G.\ 1995, \mnras, 276, 663
\bibitem[Bildfell et al.(2008)]{bildfell08} Bildfell, C., Hoekstra, H., Babul, A., \& Mahdavi, A.\ 2008, \mnras, 389, 1637 
\bibitem[Blanton et al.(2009)]{blanton09} Blanton, E.~L., Randall, S.~W., Douglass, E.~M., Sarazin, C.~L., Clarke, T.~E., \& McNamara, B.~R.\ 2009, \apjl, 697, L95 
\bibitem[Brighenti \& Mathews(1999)]{brighenti99} Brighenti, F., \& Mathews, W.~G.\ 1999, \apj, 512, 65 
\bibitem[Bruzual \& Charlot(2003)]{bc03} Bruzual, G., \& Charlot, S.\ 2003, \mnras, 344, 1000 
\bibitem[Cardiel et al.(1998)]{cardiel98} Cardiel, N., Gorgas, J., \& Aragon-Salamanca, A.\ 1998, \mnras, 298, 977 
\bibitem[Cid Fernandes et al.(2001)]{cidfernandes01} Cid Fernandes, R., Heckman, T., Schmitt, H., Delgado, R.~M.~G., \& Storchi-Bergmann, T.\ 2001, \apj, 558, 81 
\bibitem[Ciotti \& Ostriker(2007)]{ciotti07} Ciotti, L., \& Ostriker, J.~P.\ 2007, \apj, 665, 1038
\bibitem[Ciotti et al.(2009a)]{ciotti09a} Ciotti, L., Ostriker, J.~P., \& Proga, D.\ 2009a, \apj, 699, 89 (Paper I)
\bibitem[Ciotti et al.(2010)]{ciotti10} Ciotti, L., Ostriker, J.~P., \& Proga, D.\ 2010, \apj, 717, 708 (Paper III)
\bibitem[Clarke et al.(2009)]{clarke09} Clarke, T.~E., Blanton, E.~L., Sarazin, C.~L., Anderson, L.~D., Gopal-Krishna, Douglass, E.~M., \& Kassim, N.~E.\ 2009, \apj, 697, 1481 
\bibitem[C{\^o}t{\'e} et al.(2006)]{cote06} C{\^o}t{\'e}, P., et al.\ 2006, \apjs, 165, 57 
\bibitem[Cowie \& Songaila(1977)]{cowie77} Cowie, L.~L., \& Songaila, A.\ 1977, \nat, 266, 501 
\bibitem[Decarli et al.(2007)]{decarli07} Decarli, R., Gavazzi, G., Arosio, I., Cortese, L., Boselli, A., Bonfanti, C., \& Colpi, M.\ 2007, \mnras, 381, 136 
\bibitem[D'Ercole et al.(2000)]{dercole00} D'Ercole, A., Recchi, S., \& Ciotti, L.\ 2000, \apj, 533, 799 
\bibitem[Di Matteo et al.(2008)]{dimatteo08} Di Matteo, T., Colberg, J., Springel, V., Hernquist, L., \& Sijacki, D.\ 2008, \apj, 676, 33 
\bibitem[Domainko et al.(2006)]{domainko06} Domainko, W., et al.\ 2006, \aap, 452, 795 
\bibitem[Elvis et al.(1994)]{elvis94} Elvis, M., et al.\ 1994, \apjs, 95, 1 
\bibitem[Fabian(1994)]{fabian94} Fabian, A.~C.\ 1994, \araa, 32, 277 
\bibitem[Fabian et al.(2005)]{fabian05} Fabian, A.~C., Sanders, J.~S., Taylor, G.~B., \& Allen, S.~W.\ 2005, \mnras, 360, L20 
\bibitem[Fabian et al.(2006)]{fabian06} Fabian, A.~C., Celotti, A., \& Erlund, M.~C.\ 2006, \mnras, 373, L16
\bibitem[Fabian et al.(2009)]{fabian09} Fabian, A.~C., Vasudevan, R.~V., Mushotzky, R.~F., Winter, L.~M., \& Reynolds, C.~S.\ 2009, \mnras, 394, L89 
\bibitem[Ferrarese \& Merritt(2000)]{ferrarese00} Ferrarese, L., \& Merritt, D.\ 2000, \apjl, 539, L9 
\bibitem[Ferrarese(2002)]{ferrarese02} Ferrarese, L.\ 2002, \apj, 578, 90
\bibitem[Ferrari(1998)]{ferrari98} Ferrari, A.\ 1998, \araa, 36, 539 
\bibitem[Fukazawa et al.(2006)]{fukazawa06} Fukazawa, Y., Botoya-Nonesa, J.~G., Pu, J., Ohto, A., \& Kawano, N.\ 2006, \apj, 636, 698 
\bibitem[Fukugita et al.(1996)]{fukugita96} Fukugita, M., Ichikawa, T., Gunn, J.~E., Doi, M., Shimasaku, K., \& Schneider, D.~P.\ 1996, \aj, 111, 1748 
\bibitem[Gabor et al.(2009)]{gabor09} Gabor, J.~M., et al.\ 2009, \apj, 691, 705 
\bibitem[Gaetz et al.(1987)]{gaetz87} Gaetz, T.~J., Salpeter, E.~E., \& Shaviv, G.\ 1987, \apj, 316, 530 
\bibitem[Galametz et al.(2009)]{galametz09} Galametz, A., et al.\ 2009, \apj, 694, 1309 
\bibitem[Gallo et al.(2008)]{gallo08} Gallo, E., Treu, T., Jacob, J., Woo, J.-H., Marshall, P.~J., \& Antonucci, R.\ 2008, \apj, 680, 154 
\bibitem[Gebhardt et al.(2000)]{gebhardt00} Gebhardt, K., et al.\ 2000, \apjl, 539, L13
\bibitem[Gilmour et al.(2007)]{gilmour07} Gilmour, R., Gray, M.~E., Almaini, O., Best, P., Wolf, C., Meisenheimer, K., Papovich, C., \& Bell, E.\ 2007, \mnras, 380, 1467 
\bibitem[Graham et al.(2008)]{graham08} Graham, J., Fabian, A.~C., \& Sanders, J.~S.\ 2008, \mnras, 391, 1749 
\bibitem[Grogin et al.(2005)]{grogin05} Grogin, N.~A., et al.\ 2005, \apjl, 627, L97 
\bibitem[G{\"u}ltekin et al.(2009)]{gultekin09} G{\"u}ltekin, K., et al.\ 2009, \apj, 698, 198 
\bibitem[Gunn \& Gott(1972)]{gunn72} Gunn, J.~E., \& Gott, J.~R.~I.\ 1972, \apj, 176, 1 
\bibitem[H{\"a}ring \& Rix(2004)]{haring04} H{\"a}ring, N., \& Rix, H.-W.\ 2004, \apjl, 604, L89
\bibitem[Hasinger(2008)]{hasinger08} Hasinger, G.\ 2008, \aap, 490, 905 
\bibitem[Hester(2006)]{hester06} Hester, J.~A.\ 2006, \apj, 647, 910 
\bibitem[Hickox et al.(2009)]{hickox09} Hickox, R.~C., et al.\ 2009, \apj, 696, 891 
\bibitem[Humphrey et al.(2006)]{humphrey06} Humphrey, P.~J., Buote, D.~A., Gastaldello, F., Zappacosta, L., Bullock, J.~S., Brighenti, F., \& Mathews, W.~G.\ 2006, \apj, 646, 899 
\bibitem[Jahnke et al.(2004)]{jahnke04} Jahnke, K., et al.\ 2004, \apj, 614, 568 
\bibitem[Johansson et al.(2009)]{johansson09} Johansson, P.~H., Naab, T., \& Burkert, A.\ 2009, \apj, 690, 802
\bibitem[Kapferer et al.(2009)]{kapferer09} Kapferer, W., Sluka, C., Schindler, S., Ferrari, C., \& Ziegler, B.\ 2009, \aap, 499, 87 
\bibitem[Kawata \& Mulchaey(2008)]{kawata08} Kawata, D., \& Mulchaey, J.~S.\ 2008, \apjl, 672, L103 
\bibitem[Kim et al.(2008a)]{kim08a} Kim, D.-W., Kim, E., Fabbiano, G., \& Trinchieri, G.\ 2008, \apj, 688, 931 
\bibitem[Kim et al.(2008b)]{kim08b} Kim, M., Ho, L.~C., Peng, C.~Y., Barth, A.~J., Im, M., Martini, P., \& Nelson, C.~H.\ 2008, \apj, 687, 767 
\bibitem[Kirkpatrick et al.(2009)]{kirkpatrick09} Kirkpatrick, C.~C., et al.\ 2009, \apj, 697, 867 
\bibitem[Kollmeier et al.(2006)]{kollmeier06} Kollmeier, J.~A., et al.\ 2006, \apj, 648, 128 
\bibitem[K{\"o}nigl(2006)]{konigl06} K{\"o}nigl, A.\ 2006, Memorie della Societa Astronomica Italiana, 77, 598
\bibitem[Kormendy \& Richstone(1995)]{kormendy95} Kormendy, J., \& Richstone, D.\ 1995, \araa, 33, 581
\bibitem[Koulouridis \& Plionis(2010)]{koulouridis10} Koulouridis, E., \& Plionis, M.\ 2010, preprint, arXiv:1003.0753 
\bibitem[Krongold et al.(2007)]{krongold07} Krongold, Y., Nicastro, F., Elvis, M., Brickhouse, N., Binette, L., Mathur, S., \& Jim{\'e}nez-Bail{\'o}n, E.\ 2007, \apj, 659, 1022 
\bibitem[Kurosawa et al.(2009a)]{kurosawa09a} Kurosawa, R., Proga, D., \& Nagamine, K.\ 2009, \apj, 707, 823 
\bibitem[Kurosawa \& Proga(2009b)]{kurosawa09b} Kurosawa, R., \& Proga, D.\ 2009, \apj, 693, 1929 
\bibitem[Kurosawa \& Proga(2009c)]{kurosawa09c} Kurosawa, R., \& Proga, D.\ 2009, \mnras, 397, 1791 
\bibitem[Larson et al.(1980)]{larson80} Larson, R.~B., Tinsley, B.~M., \& Caldwell, C.~N.\ 1980, \apj, 237, 692 
\bibitem[Letawe et al.(2008)]{letawe08} Letawe, Y., Magain, P., Letawe, G., Courbin, F., \& Hutsem{\'e}kers, D.\ 2008, \apj, 679, 967 
\bibitem[Li et al.(2008)]{li08} Li, C., Kauffmann, G., Heckman, T.~M., White, S.~D.~M., \& Jing, Y.~P.\ 2008, \mnras, 385, 1915 
\bibitem[Livio et al.(1980)]{livio80} Livio, M., Regev, O., \& Shaviv, G.\ 1980, \apjl, 240, L83 
\bibitem[Lucero \& Young(2007)]{lucero07} Lucero, D.~M., \& Young, L.~M.\ 2007, \aj, 134, 2148 
\bibitem[Magorrian et al.(1998)]{magorrian98} Magorrian, J., et al.\ 1998, \aj, 115, 2285
\bibitem[Marconi \& Hunt(2003)]{marconi03} Marconi, A., \& Hunt, L.~K.\ 2003, \apjl, 589, L21
\bibitem[Martini(2004)]{martini04} Martini, P.\ 2004, Coevolution of Black Holes and Galaxies, 169
\bibitem[Martini et al.(2009)]{martini09} Martini, P., Sivakoff, G.~R., \& Mulchaey, J.~S.\ 2009, \apj, 701, 66 
\bibitem[Mart{\'{\i}}nez et al.(2008)]{martinez08} Mart{\'{\i}}nez, H.~J., Coenda, V., \& Muriel, H.\ 2008, \mnras, 391, 585 
\bibitem[McNamara \& Nulsen(2007)]{mcnamara07} McNamara, B.~R., \& Nulsen, P.~E.~J.\ 2007, \araa, 45, 117 
\bibitem[McCarthy et al.(2008)]{mccarthy08} McCarthy, I.~G., Frenk, C.~S., Font, A.~S., Lacey, C.~G., Bower, R.~G., Mitchell, N.~L., Balogh, M.~L., \& Theuns, T.\ 2008, \mnras, 383, 593 
\bibitem[Merritt(1983)]{merritt83} Merritt, D.\ 1983, \apj, 264, 24 
\bibitem[Moore et al.(1999)]{moore99} Moore, B., Lake, G., Quinn, T., \& Stadel, J.\ 1999, \mnras, 304, 465 
\bibitem[Nagamine et al.(2006)]{nagamine06} Nagamine, K., Ostriker, J.~P., Fukugita, M., \& Cen, R.\ 2006, \apj, 653, 881 
\bibitem[Narayan et al.(1996)]{narayan96} Narayan, R., McClintock, J.~E., \& Yi, I.\ 1996, \apj, 457, 821 
\bibitem[Nepveu(1985)]{nepveu85} Nepveu, M.\ 1985, \aap, 149, 459 
\bibitem[Nulsen(1982)]{nulsen82} Nulsen, P.~E.~J.\ 1982, \mnras, 198, 1007 
\bibitem[O'Dea et al.(2008)]{odea08} O'Dea, C.~P., et al.\ 2008, \apj, 681, 1035 
\bibitem[O'Sullivan et al.(2005)]{osullivan05} O'Sullivan, E., Vrtilek, J.~M., \& Kempner, J.~C.\ 2005, \apjl, 624, L77 
\bibitem[Owen et al.(2000)]{owen00} Owen, F.~N., Eilek, J.~A., \& Kassim, N.~E.\ 2000, \apj, 543, 611 
\bibitem[Pagel(2009)]{pagel09} Pagel, B.~E.~J.\ 2009, Nucleosynthesis and Chemical Evolution of Galaxies, 2nd ed., Cambridge University Press
\bibitem[Pei(1992)]{pei92} Pei, Y.~C.\ 1992, \apj, 395, 130 
\bibitem[Pellegrini et al.(2009)]{pellegrini09} Pellegrini, S., Ciotti, L., \& Ostriker, J.~P.\ 2009, Advances in Space Research, 44, 340 
\bibitem[Pipino et al.(2009)]{pipino09} Pipino, A., Kaviraj, S., Bildfell, C., Babul, A., Hoekstra, H., \& Silk, J.\ 2009, \mnras, 395, 462 
\bibitem[Quilis et al.(2000)]{quilis00} Quilis, V., Moore, B., \& Bower, R.\ 2000, Science, 288, 1617 
\bibitem[Rafferty et al.(2006)]{rafferty06} Rafferty, D.~A., McNamara, B.~R., Nulsen, P.~E.~J., \& Wise, M.~W.\ 2006, \apj, 652, 216 
\bibitem[Rafferty et al.(2008)]{rafferty08} Rafferty, D.~A., McNamara, B.~R., \& Nulsen, P.~E.~J.\ 2008, \apj, 687, 899 
\bibitem[Randall et al.(2008)]{randall08} Randall, S., Nulsen, P., Forman, W.~R., Jones, C., Machacek, M., Murray, S.~S., \& Maughan, B.\ 2008, \apj, 688, 208 
\bibitem[Rasmussen \& Ponman(2009)]{rasmussen09} Rasmussen, J., \& Ponman, T.~J.\ 2009, \mnras, 399, 239 
\bibitem[Reichard et al.(2009)]{reichard09} Reichard, T.~A., Heckman, T.~M., Rudnick, G., Brinchmann, J., Kauffmann, G., \& Wild, V.\ 2009, \apj, 691, 1005 
\bibitem[Reiprich et al.(2004)]{reiprich04} Reiprich, T., Kempner, J., \& Soker, N.\ 2004, The Riddle of Cooling Flows in Galaxies and Clusters of galaxies
\bibitem[Renzini(2006)]{renzini06} Renzini, A.\ 2006, \araa, 44, 141 
\bibitem[Roediger \& Br{\"u}ggen(2008)]{roediger08} Roediger, E., \& Br{\"u}ggen, M.\ 2008, \mnras, 388, L89 
\bibitem[Salom{\'e} et al.(2006)]{salome06} Salom{\'e}, P., et al.\ 2006, \aap, 454, 437 
\bibitem[S{\'a}nchez et al.(2004)]{sanchez04} S{\'a}nchez, S.~F., et al.\ 2004, \apj, 614, 586 
\bibitem[Sazonov et al.(2004)]{sazonov04} Sazonov, S.~Y., Ostriker, J.~P., \& Sunyaev, R.~A.\ 2004, \mnras, 347, 144
\bibitem[Sazonov et al.(2005)]{sazonov05} Sazonov, S.~Y., Ostriker, J.~P., Ciotti, L., \& Sunyaev, R.~A.\ 2005, \mnras, 358, 168
\bibitem[Schawinski et al.(2009)]{schawinski09} Schawinski, K., Virani, S., Simmons, B., Urry, C.~M., Treister, E., Kaviraj, S., \& Kushkuley, B.\ 2009, \apjl, 692, L19 
\bibitem[Schindler \& Diaferio(2008)]{schindler08} Schindler, S., \& Diaferio, A.\ 2008, Space Science Reviews, 134, 363 
\bibitem[Shi et al.(2009)]{shi09} Shi, Y., Rieke, G.~H., Ogle, P., Jiang, L., \& Diamond-Stanic, A.~M.\ 2009, \apj, 703, 1107 
\bibitem[Shin et al.(2010)]{shin10} Shin, M.-S., Ostriker, J.~P., \& Ciotti, L.\ 2010, \apj, 711, 268 (Paper II)
\bibitem[Shin et al.(2011)]{shin11} Shin, M.-S., Strauss, M.~A., \& Tojeiro, R.\ 2011, \mnras, 410, 1583
\bibitem[Silverman et al.(2009)]{silverman09} Silverman, J.~D., et al.\ 2009, \apj, 696, 396 
\bibitem[Sivanandam et al.(2009)]{sivanandam09} Sivanandam, S., Zabludoff, A.~I., Zaritsky, D., Gonzalez, A.~H., \& Kelson, D.~D.\ 2009, \apj, 691, 1787 
\bibitem[Smith et al.(2009)]{smith09} Smith, R.~J., Lucey, J.~R., Hudson, M.~J., Allanson, S.~P., Bridges, T.~J., Hornschemeier, A.~E., Marzke, R.~O., \& Miller, N.~A.\ 2009, \mnras, 392, 1265 
\bibitem[Soltan(1982)]{soltan82} Soltan, A.\ 1982, \mnras, 200, 115
\bibitem[Spitzer \& Baade(1951)]{spitzer51} Spitzer, L.~J., \& Baade, W.\ 1951, \apj, 113, 413
\bibitem[Springel et al.(2005a)]{springel05a} Springel, V., Di Matteo, T., \& Hernquist, L.\ 2005, \apjl, 620, L79
\bibitem[Stevens et al.(1999)]{stevens99} Stevens, I.~R., Acreman, D.~M., \& Ponman, T.~J.\ 1999, \mnras, 310, 663 
\bibitem[Sun et al.(2005)]{sun05} Sun, M., Vikhlinin, A., Forman, W., Jones, C., \& Murray, S.~S.\ 2005, \apj, 619, 169 
\bibitem[Tabor \& Binney(1993)]{tabor93} Tabor, G., \& Binney, J.\ 1993, \mnras, 263, 323
\bibitem[Takeda et al.(1984)]{takeda84} Takeda, H., Nulsen, P.~E.~J., \& Fabian, A.~C.\ 1984, \mnras, 208, 261 
\bibitem[Toniazzo \& Schindler(2001)]{toniazzo01} Toniazzo, T., \& Schindler, S.\ 2001, \mnras, 325, 509 
\bibitem[Tonnesen \& Bryan(2008)]{tonnesen08} Tonnesen, S., \& Bryan, G.~L.\ 2008, \apjl, 684, L9 
\bibitem[Tortora et al.(2009)]{tortora09} Tortora, C., Napolitano, N.~R., Romanowsky, A.~J., Capaccioli, M., \& Covone, G.\ 2009, \mnras, 396, 1132 
\bibitem[Valluri \& Jog(1990)]{valluri90} Valluri, M., \& Jog, C.~J.\ 1990, \apj, 357, 367
\bibitem[Vanden Berk et al.(2001)]{vandenberk01} Vanden Berk, D.~E., et al.\ 2001, \aj, 122, 549 
\bibitem[van den Bosch et al.(2008)]{vandenbosch08} van den Bosch, F.~C., Aquino, D., Yang, X., Mo, H.~J., Pasquali, A., McIntosh, D.~H., Weinmann, S.~M., \& Kang, X.\ 2008, \mnras, 387, 79 
\bibitem[van Dokkum et al.(2010)]{vandokkum10} van Dokkum, P.~G., et al.\ 2010, \apj, 709, 1018 
\bibitem[Veilleux et al.(2005)]{veilleux05} Veilleux, S., Cecil, G., \& Bland-Hawthorn, J.\ 2005, \araa, 43, 769
\bibitem[Vollmer et al.(2001)]{vollmer01} Vollmer, B., Cayatte, V., Balkowski, C., \& Duschl, W.~J.\ 2001, \apj, 561, 708 
\bibitem[Weinmann et al.(2009)]{weinmann09} Weinmann, S.~M., Kauffmann, G., van den Bosch, F.~C., Pasquali, A., McIntosh, D.~H., Mo, H., Yang, X., \& Guo, Y.\ 2009, \mnras, 394, 1213 
\bibitem[White et al.(1991)]{white91} White, D.~A., Fabian, A.~C., Forman, W., Jones, C., \& Stern, C.\ 1991, \apj, 375, 35 
\bibitem[Wilman et al.(2009)]{wilman09} Wilman, R.~J., Edge, A.~C., \& Swinbank, A.~M.\ 2009, \mnras, 395, 1355 
\bibitem[Young et al.(2009)]{young09} Young, L.~M., Bendo, G.~J., \& Lucero, D.~M.\ 2009, \aj, 137, 3053 
\bibitem[Yu \& Tremaine(2002)]{yu02} Yu, Q., \& Tremaine, S.\ 2002, \mnras, 335, 965
\end{thebibliography}
\end{document}